\documentclass[preprint,amsmath,amssymb]{revtex4}
\usepackage{pgfplots}
\usepackage{booktabs}
\usepackage{graphicx}
\usepackage{dcolumn}
\usepackage{bm}
\usepackage[skip=2pt,font=footnotesize]{caption}
\usepackage{amsmath}
\usepackage{mathrsfs}
\usepackage[english]{babel}
\usepackage{float}
\usepackage{amsfonts}
\usepackage{amssymb}
\usepackage{dsfont}
\usepackage{bbm}
\usepackage[utf8]{inputenc}
\usepackage[overload]{empheq}
\usepackage{cases}

\begin{document}
\normalsize
\title{Energy-Momentum tensor for Casimir effect of conducting plates in curved spacetime\\}
\author{Borzoo Nazari$^{b}$  \footnote{borzoo.nazari@ut.ac.ir}
 }
\affiliation{$^{b}$College of Engineering, University of Tehran, Tehran, Iran}

\begin{abstract}
Brown and Maclay \cite{Brown} found the energy-momentum tensor for the Casimir effect of parallel plates in 1969. We find its curved spacetime version in a static background using the point splitting regularization method. Previous results in the literature are repeated and some consequences discussed.

\end{abstract}
\maketitle
\section{Introduction}
Some aspects of the final theory of quantum gravity may reveal themselves through a probable asymptotic theory such as quantum field theory in curved spacetime. The theory has predicted important phenomena such as Hawking radiation \cite{Hawking} and particle production in expanding universe \cite{Parker} as well as the Casimir effect and radiation from accelerating conductors. The energy-momentum tensor, occupies a crucial and central role in semi-classical approach to the theory of gravity \cite{DeWitt, Birrell}.

On this ground, and after derivation of the energy momentum tensor for the Casimir effect of parallel plates in flat spacetime by Brwon and Maclay \cite{Brown}, the Casimir energy in curved spacetime has been studied by many authors investigating some physical notions such as weak principle of equivalence \cite{Calloni,Milton,Fulling,Milton2,Shajesh}, quantum vacuum structure \cite{NouriNazari,NouriNazari2} and the question that whether the vacuum energy is responsible for the cosmological constant problem or not \cite{Saharian}? Some studies have been devoted to calculating the Casimir energy in a classical background \cite{Bezerra,SorgeNew,Muniz1,Muniz2,Sorge,Sorge2019,Lima,Blasone,Buoninfante,Lambiase,BorzooEPJC,BorzooCQG,Teo2,Saharian1,Saharian2,odintsov1} while few others concerning the full energy-momentum tensor \cite{Bimonte,Bimonte1,Bimonte2,Esposito,Napolitano}.

In Ref.\cite{Esposito}, the energy-momentum tensor has been derived and analysed in Fermi coordinates for a massless scalar field confined between two conducting parallel plates in the case of weak gravitational field.  However, as is well-known and indicated in Ref.\cite{Bimonte2}, there is no mathematically essential difference between Rindler spacetime and the Fermi coordinates in curved spacetime if we neglect curvature effects (see also eq.(13.73) in Ref.\cite{Misner}). In this paper, we find $T_{\mu\nu}$ in a general static curved spacetime. Although it is a hard and tricky computation, it can be more efficiently presented using the point splitting method \cite{Christensen, Birrell}. 

In section II, after defining our set up for the Casimir apparatus, we find the Green function using a method different from the one usually used by others. We find it more simply by employing the WKB method with the aid of a known theorem in the general theory of differential equations. In section III we compute the energy-momentum tensor using the point-splitting method. Then, the Casimir energy and force will be found. Taking advantage of the Wick rotation, we find the explicit type of the divergences. To check out consistency of the computations, we reinforce the previous results in the literature. Covariant conservation of the obtained $T_{\mu\nu}$ is examined.
A careful analysis of conformal invariance and trace anomaly is done in section IV. In section V, we provide some examples in support of the developed method. The final section is devoted to some discussions.
\section{The Green function}
The apparatus is a system of two parallel plates separated by a small distance $a$ and located at distance $R$ from the source of the gravitational field. The scalar field is massless and arbitrarily coupled to gravity with Dirichlet boundary condition on plates. The spacetime metric is assumed to be
\begin{eqnarray}\label{eq1}
ds^2 = (1 +2\gamma_0 +2 \lambda_0 z) dt^2- (1+2\gamma_1 +2 \lambda_1 z)\left(dx^2+dy^2+dz^2\right),
\end{eqnarray}
Our motivation for using this type of metric is related to the fact that a typical gravitational potential can be expanded, up to second order perturbation, in the space between the plates as $Gm/c^2r=1+2\gamma_0+2\lambda_0 z+...$ where $\gamma_0=-\gamma_1=-\frac{Gm}{c^2R}<<1, \; \lambda_0 z=-\lambda_1 z=\frac{Gm}{c^2R^2}z<<1$ \cite{Sorge}. Hereafter we assume $c=\hbar=1$.

To regularize the energy-momentum tensor, we use the point splitting method. The energy-momentum tensor can be written according to the Hadamard two-point function which is related to the Feynman Green function by
\begin{eqnarray}\label{eq2}
H(x,\acute{x})=<[\phi(x),\phi(x^\prime)]_+>=2Im\; G_{F}(x,\acute{x})
\end{eqnarray}
The Feynman Green function satisfies
\begin{eqnarray}\label{eq3}
(\square-\xi R)G_{F}(x,\acute{x})=-\frac{\delta(x,\acute{x})}{\sqrt{-g}}.
\end{eqnarray}
Some calculations show that the Ricci scalar $R \approx O(\gamma_i^2),\; i=0,1$, hence should be neglected by now as we will compute everything up to second order perturbation only. This does not delete $\xi$ in the next calculations since it still presents in the energy momentum tensor. Thus we have
\begin{eqnarray}\label{eq4}
\partial_{\mu}(\sqrt{-g}g^{\mu\nu}\partial_{\nu}G_{F}(x,\acute{x}))=-\delta(x,\acute{x}).
\end{eqnarray}
The planar symmetry of the apparatus in the directions $x$ and $y$ makes it easier to work with the reduced Green function $\mathbf{g}_F(z,\acute{z})$ defined by \cite{Bimonte}
\begin{eqnarray}\label{eq5}
\begin{split}
G_{F}(x,\acute{x}) &= \int \frac{d\omega dk_{\perp}}{(2\pi)^{3}} \mathbf{g}_F(z,\acute{z})e^{-i\omega(t-\acute{t})+\vec{k}_{\perp}.(\vec{x}-\acute{\vec{x}})} \\
&=\int \frac{d\omega dk_{\perp}}{(2\pi)^{3}} U(x,y,z,t;\omega,\overrightarrow{k}),
\end{split}
\end{eqnarray}
where
\begin{eqnarray}\label{eq6}
U\equiv \mathbf{g}_F(z,\acute{z})e^{-i\omega(t-\acute{t})+i\vec{k}_{\perp}.(\vec{x}-\acute{\vec{x}})},
\end{eqnarray}
and a Feynman contour is chosen in integration. Using the same relation as (\ref{eq5}) for $\delta(x,\acute{x})$ and expanding the left side of (\ref{eq4}) we find
\begin{eqnarray}\label{eq7}
\sqrt{-g}g^{11}\partial_z^2 \mathbf{g}_F(z,\acute{z})+\partial_z(\sqrt{-g}g^{11})\partial_z \mathbf{g}_F(z,\acute{z}) -\sqrt{-g}g^{11}(k_\perp^2+\frac{g_{11}}{g_{00}}\omega^2)\mathbf{g}_F(z,\acute{z})=-\delta(z-z^\prime).
\end{eqnarray}

Here we do not use the iterative procedure to find $\mathbf{g}(z,\acute{z})$, the perturbation method used in Ref.\cite{Bimonte}. Instead, we use the general theory of differential equations and the following theorem \cite{Wylie}:

\textbf{Theorem $1$:}
\emph{The Green function for the differential equation} \noindent
\begin{eqnarray}\label{eq8}
p_0(z)y''(z)+p_1(z) y'(z)+p_2y(z)=f(z,\acute{z}),
\end{eqnarray}

\emph{which is defined on the interval $[a,b]$, along with the boundary conditions}
\begin{eqnarray}\label{eq9}
\alpha_1 y(a)=\alpha_2 \partial_zy(a), \; \beta_1 y(b)=\beta_2 \partial_zy(b),
\end{eqnarray}
\emph{is given by}
\begin{subequations}
\begin{align}[left = {G(z,\acute{z})=\empheqlbrace\,}\,]
    &\frac{Y_1(z)Y_2(\acute{z})}{W(\acute{z})p_0(\acute{z})} \;\;\;\;\;\;\;\;\;\;\;\;\;\;\;\; z<\acute{z},  \label{eq10a} \\ 
    &\frac{Y_1(\acute{z})Y_2(z)}{W(\acute{z})p_0(\acute{z})} \;\;\;\;\;\;\;\;\;\;\;\;\;\;\;\;z>\acute{z},  \label{eq10b}
\end{align}
\end{subequations}
\emph{in which $Y_1(z)$ and $Y_2(z)$ are two independent solutions of the corresponding homogeneous differential equation
and $W(z)$ is the Wronskian of $Y_1(z),\;Y_2(z)$.}

To find $Y_1(z)$ and $Y_2(z)$ we use the general solution to the homogenous part of (\ref{eq6}) (see Ref.\cite{BorzooCQG} equation (15))
\begin{eqnarray}\label{eq11}
Y(z)=D_0\left(1-(\frac{\lambda}{2}+\frac{a}{4b})z\right)\sin\left(\sqrt{b}z(1+\frac{a}{4b}z)+\Theta_0 \right),
\end{eqnarray}
in which $\Theta_0$ and $D_0$ are arbitrary constants to be determined by imposing the boundary conditions and
\begin{subequations}
\begin{align}
    &a=-2B\omega^2, \;\; b=(1-2A)\omega^2-k_\perp^2, \; \label{eq12a} \\
    &A = \gamma_0-\gamma_1, \;\; B = \lambda_0-\lambda_1, \;\; \lambda\equiv \lambda_1+\lambda_0.   \label{eq12b}
\end{align}
\end{subequations}

The Dirichlet boundary condition is given by
\begin{eqnarray}\label{eq13}
G_F(z,z^\prime)|_{z=0,l}=0.
\end{eqnarray}
According to (\ref{eq5}), (\ref{eq9}) and (\ref{eq10a})-(\ref{eq10b}) this boundary condition is equivalent to
\begin{eqnarray}\label{eq14}
Y_1(0)=0, Y_2(l)=0.
\end{eqnarray}
Therefore, it is found that
\begin{subequations}
\begin{align}
    &Y_1(z)=\left(1-(\frac{\lambda}{2}+\frac{a}{4b})z\right)\sin \sqrt{b} \left(z+\frac{a}{4b}z^2\right) \;\;\;\;\;\;\;\;\;\;\;\;\;\;\;\; z<\acute{z},  \label{eq15a} \\ 
    &Y_2(z)=\left(1-(\frac{\lambda}{2}+\frac{a}{4b})z\right)\sin \sqrt{b}\left((z-l)+\frac{a}{4b}(z^2-l^2)\right) \;\;\;\;\;\;\;\;\;\;\;\;\;\;\;\;z>\acute{z}.   \label{eq15b}
\end{align}
\end{subequations}
A computation shows that the Wronskian for $Y_1$ and $Y_2$ is given by
\begin{eqnarray}\label{eq16}
W(z^\prime)=(1-\lambda z^\prime)\sqrt{b} \sin \sqrt{b}\left(l+\frac{a}{4b}l^2\right).
\end{eqnarray}
Using $p_0(z')=\sqrt{-g}g^{11}\simeq -(1+\gamma_0+\gamma_1+\lambda z')$ we arrive at

\begin{subequations}
\begin{align}[left = {\mathbf{g}_F(z,\acute{z})=-\frac{(1-\gamma_0-\gamma_1)(1-\epsilon(z+z'))}{\sqrt{b} \sin \sqrt{b}\left(l+\frac{a}{4b}l^2\right)}\empheqlbrace\,}\,]
     & \sin  \left(\sqrt{b}z+\frac{a}{4\sqrt{b}}z^2\right) \times \nonumber \\
      &\sin \left(\sqrt{b}(\acute{z}-l)+\frac{a}{4\sqrt{b}}(\acute{z}^2-l^2)\right)\;\;\;\;\;\;\;\;\;\;\;\;\;\;\;\; z<\acute{z},  \label{eq17a} \\
       \nonumber \\ &\sin  \left(\sqrt{b}\acute{z}+\frac{a}{4\sqrt{b}}\acute{z}^2\right) \times \nonumber \\
       &\sin \left(\sqrt{b}(z-l)+\frac{a}{4\sqrt{b}}(z^2-l^2)\right) \;\;\;\;\;\;\;\;\;\;\;\;\;\;\;\;\acute{z}<z.   \label{eq17b}
\end{align}
\end{subequations}
Notice that the above green function is symmetric due to the fact that the differential equation (\ref{eq7}) is self-adjoint. In fact, the sufficient condition for (\ref{eq8}) to be self-adjoint is that $\frac{dp_0(z)}{dz}=p_1(z)$ \cite{Hassani}.

By expanding in terms of $a$, we finally find the green function up to second order perturbation as follows:
\begin{subequations}
\begin{align}
\mathbf{z<\acute{z}}: \nonumber \\
\mathbf{g}_F(z,\acute{z})=&\frac{1-\gamma_0-\gamma_1-\lambda(z+z')}{2\sqrt{b} \sin \sqrt{b}\left(l+\frac{a}{4b}l^2\right)} \left\{cos(\sqrt{b}\alpha)-cos(\sqrt{b}\beta) + \right. \nonumber \\
&\left. \frac{a}{4\sqrt{b}}\left((z^2-\acute{z}^2+l^2)sin(\sqrt{b}\beta)-(z^2+\acute{z}^2-l^2)sin(\sqrt{b}\alpha)\right)  \right\},  \label{eq18a} \\
\alpha=z+\acute{z}&-l, \; \beta=z-\acute{z}+l=\Delta z+l.  \label{eq18b}
\end{align}
\end{subequations}
To find $\mathbf{g}_F(z,\acute{z})$ for $z>z'$ it suffices to do the interchange $z\leftrightarrow z'$ as the green function is symmetric.

\section{The energy-momentum tensor}
The classical energy-momentum tensor of a scalar field in an arbitrary $n$-dimenstional spacetime is given by \cite{Birrell}:
\begin{eqnarray}\label{eq19}
T_{\mu \nu} &= (1-2\xi)\phi_{;\mu}\phi_{;\nu} +(2\xi-\frac{1}{2})g_{\mu\nu} \phi^{;\lambda}\phi_{;\lambda}-2\xi \phi_{;\mu\nu}\phi+\frac{2}{n}\xi g_{\mu\nu} \phi \Box \phi \\ \nonumber
&-\xi ( G_{\mu\nu} +\frac{2(n-1)}{n}\xi R g_{\mu\nu})\phi^2+ 2[ \frac{1}{4}-(1-\frac{1}{n})\xi]m^2 g_{\mu\nu} \phi^2,
\end{eqnarray}
in which $\Box=g^{\alpha\beta} \phi_{;\alpha\beta}$ and $G_{\mu\nu}$ is the Einstein tensor. As is commonly known, the expectation value of this energy-momentum tensor is divergent when evaluated at a typical point in curved spacetime. In fact, this is a typical behavior of the problems consisting of taking the expectation value of the operators quadratic in terms of the filed strength \cite{Birrell}.

One can see \cite{Christensen} that after employing the point splitting method the enenergy-momentum tensor takes the form
\begin{eqnarray} \label{eq20}
\langle T_{\mu \nu} \rangle &=& \lim_{x' \to x}\biggr[
{(1-2\xi)\over 4}\Bigr(G^{(1)}_{\; ;\mu' \nu}+G^{(1)}_{\; ;\mu \nu'}\Bigr)
+\left(\xi -{1\over 4} \right)g_{\mu \nu}
G^{(1)  \; \; \sigma'}_{\; ; \; \sigma} \nonumber \\
&-& {\xi \over 2} \Bigr(G^{(1)}_{\; ; \mu \nu}+G^{(1)}_{\; ;\mu' \nu'} \Bigr)
+{\xi \over 8}g_{\mu \nu}\Bigr(G^{(1) \; \; \sigma}_{\; ; \; \sigma}
+G^{(1) \; \; \sigma'}_{\; ;\;\sigma'} \Bigr)
+{\xi \over 2}G_{\mu \nu}G^{(1)} \nonumber \\
&+&\frac{3}{4}\xi^2 R g_{\mu \nu}G^{(1)}+\frac{3\xi-1}{4}m^2g_{\mu \nu}G^{(1)} \biggr].
\end{eqnarray}
in which
\begin{equation}\label{eq21}
G^{(1)}(x,x')= \langle [\phi(x),\phi(x')]_{+} \rangle = 2Im \; G_{F},
\end{equation}
is the Hadamard function and $;\mu'$ denotes differentiation with respect to $x^\prime$. 

The main idea behind the point splitting (point-separation) method is that we avoid taking the above limit by separating the points using the bivector $P$ which is responsible for the parallel transport of any tensor field from point $x$ to another distinct point $x^\prime$ along a geodesy which connects $x$ to $x^\prime$. Thus, for instance, we do the replacements such as
\begin{equation}\label{eq22}
\lim_{x' \to x} G^{(1)}_{\; ;\mu' \nu} \rightarrow P^{\mu^\prime}_\mu G^{(1)}_{\; ;\mu' \nu},
\end{equation}
in order to carry out the above limiting process.
After the calculations done, $P^{\mu^\prime}_\mu$ will be replaced by unity, i.e. by $\delta^\mu_\mu$.

The bivector $P^{\nu^\prime}_\mu$ is given by \cite{Esposito}
\begin{equation}\label{eq23}
P_{\; \nu'}^{\mu}=g^{\mu \rho} \eta_{ab} e_{\; \rho}^{b}
e_{\; \nu'}^{a}.
\end{equation}
The normalized vielbeins $e_{\; \rho}^{b}$ for the metric (\ref{eq1}) are
\begin{equation}\label{eq24}
e_{\; \mu}^{0}=\sqrt{g_{00}} \delta_{\; \mu}^{0}, \;
e_{\; \mu}^{i}=\sqrt{|g_{11}|}\delta_{\; \mu}^{i} \; \; i=1,2,3.
\end{equation}
Therefore, we find
\begin{equation}\label{eq25}
P_{\; \nu'}^{\mu}=(\sqrt{\frac{g_{0'0'}}{g_{00}}}, \sqrt{\frac{g_{1'1'}}{g_{11}}}, \sqrt{\frac{g_{1'1'}}{g_{11}}},\sqrt{\frac{g_{1'1'}}{g_{11}}}).
\end{equation}
Now, we return to (\ref{eq20}). First, note that the last term in (\ref{eq20}) which contains $G_{\mu\nu}$ vanishes for the metric (\ref{eq1}) as it is a second order term. By using (\ref{eq5}) and the replacements like (\ref{eq22}) we rewrite (\ref{eq20}) for the massless case as

\begin{eqnarray}\label{eq26}
\langle T_{\mu \nu} \rangle &=& \frac{1}{6} Im \; \int \frac{d\omega dk_{\perp}}{(2\pi)^{3}} \biggr[
2\Bigr(P^{\mu'}_{\mu}U_{\; ;\mu' \nu}+P^{\nu'}_{\nu}U_{\; ;\mu \nu'}\Bigr)
-g_{\mu \nu}P^{\sigma}_{\sigma'}U^{  \; \; \sigma'}_{;\sigma} \nonumber \\
&-&\Bigr(U_{\; ; \mu \nu}+P^{\mu'}_{\mu}P^{\nu'}_{\nu}U_{\; ;\mu' \nu'} \Bigr)
+\frac{1}{4}g_{\mu \nu}\Bigr(U^{\; \; \sigma}_{;\sigma}
+P^{\sigma'}_{\sigma}P^{\sigma}_{\sigma'}U^{ \; \; \sigma'}_{;\sigma'} \Bigr)
\biggr] \nonumber \\
&+&(\xi-\frac{1}{6}) Im \; \int \frac{d\omega dk_{\perp}}{(2\pi)^{3}} \biggr[
-\Bigr(P^{\mu'}_{\mu}U_{\; ;\mu' \nu}+P^{\nu'}_{\nu}U_{\; ;\mu \nu'}\Bigr)
+2g_{\mu \nu} P^{\sigma}_{\sigma'}U^{  \; \; \sigma'}_{;\sigma} \nonumber \\
&-&\Bigr(U_{\; ; \mu \nu}+P^{\mu'}_{\mu}P^{\nu'}_{\nu}U_{\; ;\mu' \nu'} \Bigr)
+\frac{1}{4}g_{\mu \nu}\Bigr(U^{\; \; \sigma}_{;\sigma}
+P^{\sigma'}_{\sigma}P^{\sigma}_{\sigma'}U^{ \; \; \sigma'}_{;\sigma'} \Bigr)
\biggr].
\end{eqnarray}
To find individual components of the above energy-momentum tensor we need the Cristoffel symbols associated with the metric (\ref{eq1}) as
\begin{eqnarray} \label{eq27}
\Gamma^3_{00}=\Gamma^0_{03}=\lambda_0+O(\epsilon^2),\;\;
\Gamma^1_{13}=\Gamma^2_{23}=\Gamma^3_{33}=
-\Gamma^3_{11}=-\Gamma^3_{22}=\lambda_1+O(\epsilon^2),
\end{eqnarray}
and the following relations
\begin{eqnarray} \label{eq28}
\begin{split}
&\partial_0 U=-i\omega U, \; \partial_1 U=-ik_x U, \; \partial_2 U=-ik_y U, \\
&U_{;00'}=\omega^2U, \; U_{;1'1}=k_x^2U, \; U_{;2'2}=k_y^2U, \; U_{;3'3}=\partial_z \partial_{z'}U, \\
&U_{;00}=-\omega^2U-\lambda_0 \partial_{z}U, \; U_{;0'0'}=-\omega^2U-\lambda_0 \partial_{z'}U, \\
&U_{;11}=-k_x^2U+\lambda_1 \partial_{z}U, \; U_{;1'1'}=-k_x^2U+\lambda_1 \partial_{z'}U, \\
&U_{;22}=-k_y^2U+\lambda_1 \partial_{z}U, \; U_{;2'2'}=-k_y^2U+\lambda_1 \partial_{z'}U, \\
&U_{;33}=\partial_z^2U-\lambda_1 \partial_{z}U, \; U_{;3'3'}=\partial_{z'}^2U-\lambda_1 \partial_{z'}U.
\end{split}
\end{eqnarray}

Before tending to find energy-momentum components, we need to compute scalars $P^{\sigma}_{\sigma'}U^{  \; \; \sigma'}_{;\sigma}$ and $U^{\; \; \sigma}_{;\sigma}
+P^{\sigma'}_{\sigma}P^{\sigma}_{\sigma'}U^{ \; \; \sigma'}_{;\sigma'}$ as follows.

\begin{eqnarray} \label{eq29}
\begin{split}
P^{\sigma}_{\sigma'}U^{  \; \; \sigma'}_{;\sigma}&=g^{0'0'}\sqrt{\frac{g_{0'0'}}{g_{00}}}U_{;00'}+g^{1'1'}\sqrt{\frac{g_{1'1'}}{g_{11}}}\{U_{;11'}+U_{;22'}+U_{;33'}\}\\
&=g^{00}\omega^2U+g^{11}(k_\perp^2U+\partial_z \partial_{z'}U),
\end{split}
\end{eqnarray}
in which we have used equations (\ref{eq25}) and (\ref{eq28}). The same calculation shows
\begin{eqnarray} \label{eq30}
\begin{split}
U^{\; \; \sigma}_{;\sigma}
+P^{\sigma'}_{\sigma}P^{\sigma}_{\sigma'}U^{ \; \; \sigma'}_{;\sigma'}=
&g^{00}U_{;00}+g^{11}\{U_{;11}+U_{;22}+U_{;33}\}+    \\
&g^{0'0'}U_{;0'0'}+g^{1'1'}\{U_{;1'1'}+U_{;2'2'}+U_{;3'3'}\}\\
=&-2g^{00}\omega^2U-2k_\perp^2g^{11}U-\lambda(\partial_z+\partial_{z'})U+g^{11}(\partial_z^2 +\partial_{z'}^2)U.
\end{split}
\end{eqnarray}
Note that in the above equation it is eligible to take $x \rightarrow x'$ \emph{after} the differentiation is done.
\subsection{non-diagonal components}
The non-diagonal components of the $\langle T_{\mu \nu} \rangle$ vanish. For instance, we find $\langle T_{01} \rangle$.
Since $g_{0\mu}=0$, the second and the last terms in (\ref{eq26}) vanish. For the first and third terms, we see after using (\ref{eq6}) and (\ref{eq28}) that
\begin{eqnarray} \label{eq31}
\begin{split}
&P^{\mu'}_{0}U_{\; ;\mu' 1}+P^{\nu'}_{0}U_{\; ;1 \nu'}=P^{0'}_{0}
( U_{0'1}+U_{1'0})=(-\omega k_x U-k_x \omega U )=-2k_x \omega U, \\
&U_{\; ; 01}+P^{\mu'}_{0}P^{\nu'}_{1}U_{\; ;\mu' \nu'}= 2k_x\omega U,
\end{split}
\end{eqnarray}
and arrive at
\begin{eqnarray} \label{eq32}
\begin{split}
\langle T_{01} \rangle &=\lim_{x' \to x} \; 2Im \; \int \frac{d\omega dk_{\perp}}{(2\pi)^{3}} \biggr[2k_x\omega U\biggr] \\
&=2 \;  \int_{+\infty}^{-\infty} dk_y \int_{+\infty}^{-\infty}d\omega \frac{ 1 }{(2\pi)^{3}} \int_{+\infty}^{-\infty}dk_x\biggr[2k_x\omega  Z(z,z')\biggr]=0,
\end{split}
\end{eqnarray}
since $\omega$ and $Z(z,z')$ are even functions in terms of $k_x$, hence the integrand in (\ref{eq32}) is an odd function in terms of $k_x$.

\subsection{ computation of $<T_{00}>$}
After using (\ref{eq26})-(\ref{eq30}) we find

\begin{eqnarray}\label{eq33}
\begin{split}
&\langle T_{00} \rangle = \frac{1}{6} Im \; \int \frac{d\omega dk_{\perp}}{(2\pi)^{3}} \biggr[
2\Bigr(2\omega^2\Bigr)U
-g_{00}\{g^{00}\omega^2U+g^{11}(k_\perp^2U+\partial_z \partial_{z'}U)\}   \\
&-\Bigr( -2\omega^2 U -\lambda_0 (\partial_z+\partial_{z'})U \Bigr)  \\
&+\frac{1}{4}g_{00}\Bigr(-2g^{00}\omega^2U-2k_\perp^2g^{11}U-\lambda(\partial_z+\partial_{z'})U+g^{11}(\partial_z^2 +\partial_{z'}^2)U \Bigr)
\biggr]  \\
&+(\xi-\frac{1}{6}) Im \; \int \frac{d\omega dk_{\perp}}{(2\pi)^{3}} \biggr[
-2\omega^2U
+2g_{00}\{g^{00}\omega^2U+g^{11}(k_\perp^2U+\partial_z \partial_{z'}U)\}   \\
&-\Bigr( -2\omega^2 U -\lambda_0 (\partial_z+\partial_{z'})U \Bigr)  \\
&+\frac{1}{4}g_{00}\Bigr(-2g^{00}\omega^2U-2k_\perp^2g^{11}U-\lambda(\partial_z+\partial_{z'})U+g^{11}(\partial_z^2 +\partial_{z'}^2)U \Bigr)
\biggr]
\end{split}
\end{eqnarray}
which can be written as follows
\begin{eqnarray}\label{eq34}
\begin{split}
\langle T_{00} \rangle
&=\frac{1}{6} \lim_{z' \to z} Im  \int \frac{d\omega dk_{\perp}}{(2\pi)^{3}} \biggr[
\frac{9}{2}\omega^2 - \frac{3}{2}\frac{g_{00}}{g_{11}} k_\perp^2+\frac{g_{00}}{4g_{11}}\Bigr(\partial_z^2 +\partial_{z'}^2-4\partial_z \partial_{z'}\Bigr) \\
&\;\;\;\;\;\;\;\;\;\;\;\;\;\;\;\;\;\;\;\;\;\;\;\;\;\;\;\;\;\;\;\;\;\;\;\;\;\;\;\;\;\;\;\;\;\;\;\;\;\;\;\;\;\;\;\;\;\;\;\;
+{3\lambda_0-\lambda_1\over 4}(\partial_z+\partial_{z'})
 \biggr]\mathbf{g}_F \\
&+(\xi-\frac{1}{6}) \lim_{z' \to z} Im  \int \frac{d\omega dk_{\perp}}{(2\pi)^{3}} \biggr[\frac{3}{2}\omega^2 + \frac{3}{2}\frac{g_{00}}{g_{11}} k_\perp^2+\frac{g_{00}}{4g_{11}}\Bigr((\partial_z^2 +\partial_{z'}^2)+8\partial_z \partial_{z'}\Bigr) \\
&\;\;\;\;\;\;\;\;\;\;\;\;\;\;\;\;\;\;\;\;\;\;\;\;\;\;\;\;\;\;\;\;\;\;\;\;\;\;\;\;\;\;\;\;\;\;\;\;\;\;\;\;\;\;\;\;\;\;\;\;
+{3\lambda_0-\lambda_1\over 4}(\partial_z+\partial_{z'})
 \biggr]\mathbf{g}_F.
 \end{split}
\end{eqnarray}
Before doing integrations in the above equation, we need to find $\mathbf{g}_F, \; (\partial_z+\partial_{z'})\mathbf{g}_F$ and $ \partial_z \partial_{z'}\mathbf{g}_F $ separately. After a careful calculation, we find up to second order perturbations in terms of $\lambda, \gamma$ (see Appendix A)

\begin{subequations}
\begin{align}
\mathbf{g}_F=&\frac{1}{2}(N-M), \label{eq35a} \\
M=& -(1-\gamma_0-\gamma_1)\frac{cos(\sqrt{b}\alpha)}{\sqrt{b}sin(\sqrt{b}l)}+ \frac{al^2}{4b}\frac{cos(\sqrt{b}\alpha)cos(\sqrt{b}l)}{sin^2(\sqrt{b}l)}+\nonumber  \\ &\;\;\;\;\;\;\;\;\;\;\;\;\;\;\;\;\frac{a}{4\sqrt{b}}(z^2+z'^2-l^2) \frac{sin(\sqrt{b}\alpha)}{\sqrt{b}sin(\sqrt{b}l)} +2\epsilon z \frac{cos(\sqrt{b}\alpha)}{\sqrt{b}sin(\sqrt{b}l)} \label{eq35b} \\
N=& -(1-\gamma_0-\gamma_1)\frac{cos(\sqrt{b}\beta)}{\sqrt{b}sin(\sqrt{b}l)}+ \frac{al^2}{4b}\frac{cos(\sqrt{b}\beta)cos(\sqrt{b}l)}{sin^2(\sqrt{b}l)}+ \nonumber \\
&\;\;\;\;\;\;\;\;\;\;\;\;\;\;\; \frac{a}{4\sqrt{b}}(z^2-z'^2+l^2)\frac{sin(\sqrt{b}\beta)}{\sqrt{b}sin(\sqrt{b}l)} +2\epsilon z \frac{cos(\sqrt{b}\beta)}{\sqrt{b}sin(\sqrt{b}l)}, \label{eq35c}\\
\epsilon=&\frac{\lambda}{2}+\frac{a}{4b}. \label{eq35d}
\end{align}
\end{subequations}
and
\begin{equation}\label{eq36}
(\partial_z\partial_{z'})\mathbf{g}_F=\epsilon\frac{sin(\sqrt{b}\alpha)}{sin(\sqrt{b}l)}
-\frac{az}{2\sqrt{b}}\left(\frac{cos(\sqrt{b}\alpha)}{sin(\sqrt{b}l)}+\frac{cos(\sqrt{b}\beta)}{sin(\sqrt{b}l)}\right)
+b\frac{M+N}{2},
\end{equation}

\begin{equation}\label{eq37}
(\partial^2_z+\partial^2_{z'})\mathbf{g}_F=\lambda\frac{sin(\sqrt{b}\alpha)}{sin(\sqrt{b}l)}
-\frac{az}{\sqrt{b}}\left(\frac{cos(\sqrt{b}\alpha)}{sin(\sqrt{b}l)}-\frac{cos(\sqrt{b}\beta)}{sin(\sqrt{b}l)}\right)
+b(M-N),
\end{equation}

\begin{eqnarray}\label{eq38}
\begin{split}
\lambda_1(\partial_z+\partial_{z'})\mathbf{g}_F &= \frac{-\lambda_1}{2sin(\sqrt{b}l)}(sin(\sqrt{b}\alpha)-sin(\sqrt{b}\beta))+
 \frac{-\lambda_1}{2sin(\sqrt{b}l)}(sin(\sqrt{b}\alpha)+sin(\sqrt{b}\beta))  \\
&=-\lambda_1\frac{sin(\sqrt{b}\alpha)}{sin(\sqrt{b}l)}.
\end{split}
\end{eqnarray}
Note that in our calculations we frequently use typical approximations of the kind
\begin{equation}\label{eq39}
\varepsilon f(\varepsilon,...)=\varepsilon f(0,...)+O(\varepsilon^2).
\end{equation}
For example
\begin{equation}\label{eq40}
\lambda_1 \frac{1-\epsilon(z+z')}{\sqrt{b}sin(\sqrt{b}\alpha+\lambda_0 z)}=\lambda_1 \frac{1}{\sqrt{b}sin(\sqrt{b}\alpha)}+O(\lambda_1\lambda_0)+O(\lambda_1\epsilon).
\end{equation}
To have insight into the divergent parts of equations (\ref{eq34})-(\ref{eq38}), we need to first analyse the flat spacetime case. Explicit computations will be done to clear the type of divergences. 

\subsection {flat spacetime analysis of the energy-momentum tensor}
The flat space $\langle T_{00} \rangle$ is given by letting $\lambda_0=\lambda_1=\gamma_0=\gamma_1=0$ in equation (\ref{eq33}). For sake of simplicity we analyse the case $\xi=0$, i.e.
\begin{eqnarray}\label{eq41}
\begin{split}
\langle T_{00} \rangle
=\lim_{z' \to z} \frac{1}{2} Im \; \int \frac{d\omega dk_{\perp}}{(2\pi)^{3}}
[\omega^2  +k_\perp^2+\partial_z \partial_{z'}]\mathbf{g}_F.
\end{split}
\end{eqnarray}
After using (\ref{eq35a}) and (\ref{eq36}) we find
\begin{eqnarray}\label{eq42}
\begin{split}
\langle T_{00} \rangle
=-\frac{1}{2} \lim_{z' \to z} Im \int \frac{d\omega dk_{\perp}}{(2\pi)^{3}}
\frac{\omega^2\cos(\sqrt{b}\beta)-k_\perp^2\cos(\sqrt{b}\alpha)}{\sqrt{b}\sin(\sqrt{b}l)}.
\end{split}
\end{eqnarray}
To compute this integral we use the Wick rotation technique discussed in Appendix B. Thus, we find
\begin{eqnarray}\label{eq43}
\begin{split}
\langle T_{00} \rangle
&=-\frac{1}{12\pi^2} \lim_{z' \to z} \int_{0}^{\infty} \frac{\kappa^3 d\kappa}{\sinh\kappa l}
\biggr[\cosh(\kappa\beta)+2\cosh(\kappa\alpha) \biggr]  \\
&=-\frac{1}{12\pi^2} \lim_{z' \to z} \int_{0}^{\infty} \frac{\kappa^3 d\kappa}{\sinh\kappa l}
\biggr[\cosh(\kappa (\Delta z+l))+2\cosh(\kappa(2z-l)) \biggr],
\end{split}
\end{eqnarray}
which in turn gives
\begin{eqnarray}\label{eq44}
\begin{split}
\langle T_{00} \rangle
=&-\frac{1}{12\pi^2}  \biggr[ \lim_{z' \to z}  \int_{0}^{\infty} \kappa^3 d\kappa e^{\kappa\Delta z}+ \int_{0}^{\infty}\frac{2\kappa^3 d\kappa}{e^{2\kappa l}-1}+
2\int_{0}^{\infty}\frac{\kappa^3 e^{2\kappa z} d\kappa}{e^{2\kappa l}-1}  \\
&+2\int_{0}^{\infty}\frac{\kappa^3 e^{2\kappa (a-z)} d\kappa}{e^{2\kappa l}-1}\biggr].
\end{split}
\end{eqnarray}
After using \cite{gradshtyn}
\begin{eqnarray}\label{eq45}
\int_{0}^{\infty}\frac{x^\nu e^{(\beta-\mu)x}}{e^{\beta x}-1}dx=\frac{1}{\beta^{\nu+1}}\Gamma(\nu+1)\zeta(\nu+1,\frac{\mu}{\beta}), \;\;\; Re\;\beta>0,\; Re \;\mu>0,\; Re\; \nu>1,
\end{eqnarray}
we find
\begin{eqnarray}\label{eq46}
\begin{split}
\langle T_{00} \rangle
=&\biggr[-\frac{1}{2\pi^2}   \lim_{z' \to z}  \frac{1}{(\Delta z)^4}- \frac{\pi^2}{1440 l^4}\biggr]-\frac{1}{16\pi^2 l^4}\biggr[
\zeta\left(4,1-\frac{z}{l}\right)+\zeta\left(4,\frac{z}{l}\right) \biggr],
\end{split}
\end{eqnarray}
in which
\begin{eqnarray}\label{eq47}
\zeta(m,x)=\sum_{n=o}^{\infty} \frac{1}{(n+x)^m},
\end{eqnarray}
is the Riemann's zeta function.

Notice that the first bracket in (\ref{eq46}) is originated from the $\beta$-dependent part of equation (\ref{eq42}) while the second bracket is due to the $\alpha$-dependent part. We will use this point later in next sections. As is evident from the $\beta$-dependent part, the first term diverges when $z' \to z$. This is the typical behaviour of the point splitting method and is not an special effect here \cite{Christensen}. Except for this point, the $\beta$-dependent part is the finite one.

A simple computation shows that the $\alpha$-dependent part is completely divergent. In fact,
\begin{eqnarray}\label{eq48}
\begin{split}
E_{\alpha}&=-\frac{1}{16\pi^2 \; l^4} \int_{0}^{l} \biggr[
\zeta\left(4,1-\frac{z}{l}\right)+\zeta\left(4,\frac{z}{l}\right) \biggr] dz \\
&=-\frac{1}{16\pi^2}\Biggr[ \int_{0}^{l}\frac{dz}{z^4}+\int_{0}^{l}\frac{dz}{(l-z)^4}
+\int_{0}^{l}dz
\sum_{n=1}^{\infty}\left( \frac{1}{(z+n l)^4}+\frac{1}{[(n+1)l-z]^4} \right) \Biggr]\\
&= -\frac{1}{4\pi^2}\left(\lim_{z \to 0^+}\frac{1}{z^3}+\lim_{z \to l^-}\frac{1}{(l-z)^3} \right).
\end{split}
\end{eqnarray}

Therefore, the $\alpha$-dependent part does not produce any finite contributions to the energy.

\subsection{regularization of $<T_{00}>$}
Based on the analysis presented in the previous section, the terms containing $\alpha$ are divergent at $z=0,z=l$ while the $\beta$-terms converge. Hereafter we first calculate $\beta$-terms in each case. Note also that $M$ is a totally $\alpha$-term.

For the first line of (\ref{eq34}), after using (\ref{eq35a}) and (\ref{eq36})-(\ref{eq38}) we find
\begin{eqnarray}\label{eq491}
\begin{split}
&\biggr[\frac{9}{2}\omega^2 - \frac{3}{2}\frac{g_{00}}{g_{11}} k_\perp^2+\frac{g_{00}}{4g_{11}}\Bigr(\partial_z^2 +\partial_{z'}^2-4\partial_z \partial_{z'}\Bigr)
+{3\lambda_0-\lambda_1\over 4}(\partial_z+\partial_{z'})
 \biggr]\frac{N-M}{2} \nonumber \\
&=\Biggr[
-3(1-\gamma_0-\gamma_1+\frac{1}{2} B z )\frac{\omega^2}{\sqrt{b}}\frac{\cos(\sqrt{b}\beta)}{sin(\sqrt{b}l)}
+\frac{3l^2}{4}\frac{a\omega^2}{b}\frac{\cos(\sqrt{b}\beta)\cos(\sqrt{b}l)}{sin^2(\sqrt{b}l)}
+\frac{3l^2}{4}\frac{a\omega^2}{b}\frac{\sin(\sqrt{b}\beta)}{sin(\sqrt{b}l)}               \\
&\;\;\;\;\;\;\;\;\;\;\;\;\;\;\;\;\;\;\;\;\;\;\;\;\;\;\;\;\;\;\;\;\;\;\;\;\;\;\;\;\;\;\;\;\;\;\;\;\;\;\;\;+6z\frac{\epsilon \; \omega^2}{\sqrt{b}}\frac{ \cos(\sqrt{b}\beta)}{sin(\sqrt{b}l)}
- \frac{3z}{4} \frac{a}{\sqrt{b}}\frac{cos(\sqrt{b}\beta)}{sin(\sqrt{b}l)}
\Biggr]
\end{split}
\end{eqnarray}
\begin{eqnarray}\label{eq49}
\begin{split}
&+\Biggr[
-(1+\gamma_0-3\gamma_1+2B z )\frac{\sqrt{b}\cos(\sqrt{b}\alpha)}{sin(\sqrt{b}l)}
 +\frac{l^2}{4}a\frac{\cos(\sqrt{b}\alpha)\cos(\sqrt{b}l)}{sin^2(\sqrt{b}l)} +\frac{1}{4}(2z^2-l^2)a\frac{\sin(\sqrt{b}\alpha)}{\sin(\sqrt{b}l)}      \\
&+2z\frac{\epsilon \sqrt{b} \cos(\sqrt{b}\alpha)}{\sin(\sqrt{b}l)}
+3(1-\gamma_0-\gamma_1+\frac{1}{2}B z )\frac{\omega^2}{\sqrt{b}}\frac{cos(\sqrt{b}\alpha)}{sin(\sqrt{b}l)}
-\frac{3l^2}{4}\frac{a\omega^2}{b}\frac{cos(\sqrt{b}\alpha)\cos(\sqrt{b}l)}{sin^2(\sqrt{b}l)}                  \\
&-\frac{3}{4}(2z^2-l^2)\frac{a\omega^2}{b}\frac{\sin(\sqrt{b}\alpha)}{\sin(\sqrt{b}l)}
-6z\frac{\epsilon\omega^2}{\sqrt{b}}\frac{\cos(\sqrt{b}\alpha)}{\sin(\sqrt{b}l)}
-\frac{z}{4} \frac{a}{\sqrt{b}}\frac{\cos(\sqrt{b}\alpha)}{\sin(\sqrt{b}l)}
-\frac{\lambda}{4}\frac{\sin(\sqrt{b}\alpha)}{\sin(\sqrt{b}l}  \\
&\;\;\;\;\;\;\;\;\;\;\;\;\;\;\;\;\;\;\;\;\;\;\;\;\;\;\;\;\;\;\;\;\;\;\;\;\;\;\;\;\;\;\;\;\;\;\;\;\;\;\;\;\;\;\;\;\;
+\frac{\lambda_1-3\lambda_0}{4} \frac{ \sin(\sqrt{b}\alpha)}{\sin(\sqrt{b}l)}  +\epsilon \frac{ \sin(\sqrt{b}\alpha)}{\sin(\sqrt{b}l)}
\Biggr].
\end{split}
\end{eqnarray}
In a similar manner, for the second line of (\ref{eq34}), after using (\ref{eq35b}) and (\ref{eq36})-(\ref{eq38}), we find
\begin{eqnarray}\label{eq50}
\begin{split}
&\Biggr[\frac{3}{2}\omega^2 + \frac{3}{2}\frac{g_{00}}{g_{11}} k_\perp^2+\frac{g_{00}}{4g_{11}}\Bigr((\partial_z^2 +\partial_{z'}^2)+8\partial_z \partial_{z'}\Bigr)+{(3\lambda_0-\lambda_1)\over 4}(\partial_z+\partial_{z'}) \Biggr]\frac{N-M}{2}    \\
&=\Biggr[
\frac{3}{2}Bz \frac{\omega^2}{\sqrt{b}} \frac{\cos(\sqrt{b}\beta)}{\sin(\sqrt{b}l)}
+\frac{3z}{4}\frac{a}{\sqrt{b}} \frac{\cos(\sqrt{b}\beta)}{\sin(\sqrt{b}l)}
\Biggr] \\
&+\Biggr[
-\frac{3}{2}Bz \frac{\omega^2}{\sqrt{b}} \frac{\cos(\sqrt{b}\alpha)}{\sin(\sqrt{b}l)}
+(1+\gamma_0-3\gamma_1+2Bz) \sqrt{b} \frac{\cos(\sqrt{b}\alpha)}{\sin(\sqrt{b}l)}
-\frac{l^2}{4} a \frac{\cos(\sqrt{b}\alpha)\cos(\sqrt{b}l)}{\sin^2(\sqrt{b}l)} \\
&-\frac{2z^2-l^2}{4}a\frac{\sin(\sqrt{b}\alpha)}{\sin(\sqrt{b}l)}
-2z\epsilon \sqrt{b} \frac{\cos(\sqrt{b}\alpha)}{\sin(\sqrt{b}l)}
+\frac{5}{4}z \frac{a}{\sqrt{b}} \frac{\cos(\sqrt{b}\alpha)}{\sin(\sqrt{b}l)} -\frac{1}{2} \epsilon \frac{\sin(\sqrt{b}\alpha)}{\sin(\sqrt{b}l)} \\
&-\frac{3\lambda_0-\lambda_1}{4} \frac{\sin(\sqrt{b}\alpha)}{\sin(\sqrt{b}l)}
-\frac{\lambda}{4}\frac{\sin(\sqrt{b}\alpha)}{\sin(\sqrt{b}l)}
\Biggr].
\end{split}
\end{eqnarray}
As is evident from (\ref{eq50}), the second line of (\ref{eq34}) is completely divergent at $z=0$ and $z=l$. This divergent part is absent in the case of conformal coupling of the field, i.e. for $\xi=1/6$.
Thus, after a Wick rotation and using other integrations in appendix B, we finally find
\begin{eqnarray}\label{eq51}
\begin{split}
&\langle T_{00} \rangle
=\frac{E_0}{l}  \biggr( 1+2\gamma_0-4\gamma_1+\frac{2}{5}\lambda_0(3l-z)-\frac{2}{5}\lambda_1(3l+4z) \biggr)+\frac{c_0}{(\Delta z)^4} \\
&\;\;\;\;\;\;\;+\frac{B}{90 \pi^2}\biggr[
-8zA_1(\alpha)+l^2A_2(\alpha)-(2z^2-l^2)A_3(\alpha)+5 A_4(\alpha)
 \biggr] \\
&+\frac{1}{12\pi^2}(\xi-\frac{1}{6})\biggr[ 6\left(1+2\gamma_0-4\gamma_1-2\lambda_1z\right)A_1(\alpha)+Bl^2A_2(\alpha)-(2z^2-l^2)B \; A_3(\alpha) \\ &\;\;\;\;\;\;\;\;\;\;\;\;\;\;\;\;\;\;\;\;\;\;\;\;\;\;\;-2(4\lambda_1+5\lambda_0) A_4(\alpha) \biggr],
\end{split}
\end{eqnarray}
in which

\begin{eqnarray}\label{eq521}
c_0=-\frac{1}{2\pi^2}\biggr[1+2\gamma_0-4\gamma_1-\frac{2}{5}(\lambda_0+4\lambda_1)z\biggr],
\end{eqnarray}
and
\begin{eqnarray}\label{eq52}
\begin{split}
A_1(\alpha)=&\frac{3}{8l^4}\biggr[\zeta(4,1-\frac{z}{l})+\zeta(4,\frac{z}{l})\biggr],  \\
A_2(\alpha)=&\frac{3}{4l^5} \biggr[ 2\zeta(4,2-\frac{z}{l})-2(1-\frac{z}{l})\zeta(5,2-\frac{z}{l})  \nonumber \\
&+2\zeta(4,1+\frac{z}{l})-\frac{2z}{l}\zeta(5,1+\frac{z}{l})+\zeta(5,1-\frac{z}{l})+\zeta(5,\frac{z}{l})
\biggr] , \\
A_3(\alpha)=&\frac{3}{4l^5}\biggr[\zeta(5,1-\frac{z}{l})-\zeta(5,\frac{z}{l})\biggr],  \\
A_4(\alpha)=&\frac{1}{4l^3}\biggr[\zeta(3,1-\frac{z}{l})-\zeta(3,\frac{z}{l})\biggr] , \\
\end{split}
\end{eqnarray}
and $E_0=-\pi^2/1440l^4$ is the Casimir energy in flat spacetime.
An important point should be stressed here. In equations (\ref{eq49}) and (\ref{eq50}), everywhere, we can replace $b=(1-2A)\omega^2-k_\perp^2$ by $b=\omega^2-k_\perp^2$ in view of the application of equation (\ref{eq38}). However, since the first and fifth terms in (\ref{eq49}) and the second term in (\ref{eq50}) are not proportional to $O(\lambda)$, hence the replacement is not eligible. In such terms, we can use the variable change $\omega'\rightarrow \omega(1-A)$ and send $b=(1-2A)\omega^2-k_\perp^2$ to $b_0=\omega'^2-k_\perp^2$ again. Consequently it makes an extra multiplicative factor of $1+3A$ which should be taken into account.
Another point is that all the functions $A_1,A_2,A_3$ and $A_4$ are divergent near the surfaces, i.e. at $z=0$ and $z=l$.

\subsection{computation of $<T_{11}>$ and $<T_{22}>$ }
The fact that there is horizontal symmetry in the space between the plates, the energy momentum tensor is the same for both $x$ and $y$ directions, i.e. $\langle T_{11} \rangle=\langle T_{22} \rangle$.  The same reasoning will ended up with the following relations which we use later:
\begin{eqnarray}\label{eq53}
\lim_{z' \to z} Im  \int \frac{d\omega dk_\perp}{(2\pi)^{3}} k_y^2\mathbf{g}_F=\lim_{z' \to z} Im  \int \frac{d\omega dk_\perp}{(2\pi)^{3}} k_x^2\mathbf{g}_F=\frac{1}{2}\lim_{z' \to z} Im  \int \frac{d\omega dk_\perp}{(2\pi)^{3}} k_\perp^2 \mathbf{g}_F
\end{eqnarray}
By using (\ref{eq26}) we find
\begin{eqnarray}\label{eq54}
\begin{split}
&\langle T_{11} \rangle =\langle T_{22} \rangle \\
&=\frac{1}{6} \lim_{z' \to z} Im  \int \frac{d\omega dk_{\perp}}{(2\pi)^{3}} \biggr[
\frac{3}{2}k_\perp^2-\frac{3}{2}\frac{g_{11}}{g_{00}}\omega^2 +\frac{1}{4}\Bigr(\partial_z^2 +\partial_{z'}^2-4\partial_z \partial_{z'}\Bigr)
+{\lambda_0-3\lambda_1\over 4}(\partial_z+\partial_{z'})
 \biggr]\mathbf{g}_F \\
&+(\xi-\frac{1}{6}) \lim_{z' \to z} Im \; \int \frac{d\omega dk_{\perp}}{(2\pi)^{3}} \biggr[
\frac{3}{2}k_\perp^2+\frac{3}{2}\frac{g_{11}}{g_{00}}\omega^2+\frac{1}{4}\Bigr(\partial_z^2 +\partial_{z'}^2+8\partial_z \partial_{z'}\Bigr)    \\
&\;\;\;\;\;\;\;\;\;\;\;\;\;\;\;\;\;\;\;\;\;\;\;\;\;\;\;\;\;\;\;\;\;\;\;\;\;\;\;\;\;\;\;\;\;\;\;\;\;\;\;
\;\;\;\;\;\;\;\;\;\;\;\;\;\;\;\;\;\;\;\;\;\;\;\;\;\;
+{\lambda_0-3\lambda_1\over 4}(\partial_z+\partial_{z'})
 \biggr]\mathbf{g}_F.
\end{split}
\end{eqnarray}
After the same process as for $\langle T_{00} \rangle$ we find for the first line of (\ref{eq54})
\begin{eqnarray}\label{eq551}
\begin{split}
&\biggr[
\frac{3}{2}k_\perp^2-\frac{3}{2}\frac{g_{11}}{g_{00}}\omega^2 +\frac{1}{4}\Bigr(\partial_z^2 +\partial_{z'}^2-4\partial_z \partial_{z'}\Bigr)
+{\lambda_0-3\lambda_1\over 4}(\partial_z+\partial_{z'})
 \biggr]\mathbf{g}_F\\
&=\frac{1}{4}\Biggr[
(1-3\gamma_0+\gamma_1-B z) \frac{\omega^2}{\sqrt{b}}\frac{\cos(\sqrt{b}\beta)}{\sin(\sqrt{b}l)}
+\frac{l^2}{4}\frac{a\omega^2}{b}\frac{\cos(\sqrt{b}\beta)\cos(\sqrt{b}l)}{\sin^2(\sqrt{b}l)}
+\frac{l^2}{4}\frac{a\omega^2}{b}\frac{\sin(\sqrt{b}\beta)}{\sin(\sqrt{b}l)} \\
&+2z\frac{\epsilon\omega^2}{\sqrt{b}}\frac{\cos(\sqrt{b}\beta)}{\sin(\sqrt{b}l)}
+(1-\gamma_0-\gamma_1)\sqrt{b}\frac{\cos(\sqrt{b}\beta)}{\sin(\sqrt{b}l)}
-\frac{l^2}{4}a\frac{\cos(\sqrt{b}\beta)\cos(\sqrt{b}l)}{\sin^2(\sqrt{b}l)} \\
&-\frac{l^2}{4}a\frac{\sin(\sqrt{b}\beta)}{\sin(\sqrt{b}l)}
-2z\epsilon \sqrt{b} \frac{\cos(\sqrt{b}\beta)}{\sin(\sqrt{b}l)}
+\frac{z}{2}\frac{a}{\sqrt{b}}\frac{\cos(\sqrt{b}\beta)}{sin(\sqrt{b}l)}\Biggr]  \\
&+\frac{1}{4}\Biggr[
(1-3\gamma_0+\gamma_1-B z) \frac{\omega^2}{\sqrt{b}}\frac{\cos(\sqrt{b}\alpha)}{\sin(\sqrt{b}l)}
-\frac{l^2}{4}\frac{a\omega^2}{b}\frac{\cos(\sqrt{b}\alpha)\cos(\sqrt{b}l)}{\sin^2(\sqrt{b}l)} \\
&-\frac{2z^2-l^2}{4}\frac{a\omega^2}{b}\frac{\sin(\sqrt{b}\alpha)}{\sin(\sqrt{b}l)}
-2z\frac{\epsilon\omega^2}{\sqrt{b}}\frac{\cos(\sqrt{b}\alpha)}{\sin(\sqrt{b}l)}
-\frac{1}{3}(1-\gamma_0-\gamma_1)\sqrt{b}\frac{\cos(\sqrt{b}\alpha)}{\sin(\sqrt{b}l)}         \\
&+\frac{l^2}{12}a\frac{\cos(\sqrt{b}\alpha)\cos(\sqrt{b}l)}{\sin^2(\sqrt{b}l)} +\frac{2z^2-l^2}{12}a\frac{\sin(\sqrt{b}\alpha)}{\sin(\sqrt{b}l)}
+\frac{2}{3}z\epsilon \sqrt{b} \frac{\cos(\sqrt{b}\alpha)}{\sin(\sqrt{b}l)}
+\frac{z}{6}\frac{a}{\sqrt{b}}\frac{\cos(\sqrt{b}\alpha)}{sin(\sqrt{b}l)}   \\
&\;\;\;\;\;\;\;\;\;\;\;\;\;\;\;\;\;\;\;\;\;\;\;\;-\frac{2}{3}\epsilon \frac{\sin(\sqrt{b}\alpha)}{\sin(\sqrt{b}l)}
+\frac{2}{3}\lambda_1\frac{\sin(\sqrt{b}\alpha)}{\sin(\sqrt{b}l)}
\Biggr]
\end{split}
\end{eqnarray}
The second line in equation (\ref{eq54}) is easily obtained from equation (\ref{eq50}). In fact, it is nothing but the equation (\ref{eq49}) multiplied by a factor of $\frac{g_{00}}{g_{11}}$ along with the exchange $\lambda_0\leftrightarrow \lambda_1$. Thus, again, using the integrations in the appendix B, we find

\begin{eqnarray}\label{eq56}
\begin{split}
\langle T_{11} \rangle=\langle T_{22} \rangle&=-\left( 1-2\gamma_1-\frac{2}{5}\lambda_0(2z-l)-\frac{2}{5}\lambda_1(3z+l) \right)\frac{E_0}{l}+\frac{c_2}{(\Delta z)^4}\\
&\;\;\;\;\;\;\;\;\;\;\;\;\;\;\;\;\;\;\;-\frac{B}{180\; \pi^2} \biggr[ 8zA_1(\alpha)-l^2A_2(\alpha)+(2z^2-l^2)A_3(\alpha)+5A_4(\alpha)
\biggr]\\
&\;\;\;\;+\frac{1}{12\pi^2}(\xi-\frac{1}{6})
\biggr[ -6\left(1-2\gamma_1-2\lambda_0 z\right)A_1(\alpha)-Bl^2A_2(\alpha)       \\
&\;\;\;\;\;\;\;\;\;\;\;\;\;\;\;\;\;\;\;\;\;\;\;\;\;\;\;\;\;\;\;\;\;\;\;\;\;\;\;\;\;\;\;\;\;\;\;\;\;\;
+(2z^2-l^2)BA_3(\alpha) +2(2\lambda_0+7\lambda_1)A_4(\alpha)\biggr],
\end{split}
\end{eqnarray}
where
\begin{eqnarray}\label{eq57}
c_2=\frac{1}{2\pi^2}\left( 1-2\gamma_1-\frac{2}{5}(2\lambda_0+3\lambda_1)z\right).
\end{eqnarray}

\subsection{computation of $<T_{33}>$}
Using (\ref{eq26}) and a similar process of previous subsections we find

\begin{eqnarray}\label{eq58}
\begin{split}
&\langle T_{33} \rangle=\\
&\frac{1}{6} \lim_{z' \to z} Im  \int \frac{d\omega dk_{\perp}}{(2\pi)^{3}} \biggr[
-\frac{3}{2}k_\perp^2-\frac{3}{2}\frac{g_{11}}{g_{00}}\omega^2 -\frac{3}{4}\Bigr(\partial_z^2 +\partial_{z'}^2-4\partial_z \partial_{z'}\Bigr)
+{\lambda_0+5\lambda_1\over 4}(\partial_z+\partial_{z'})
 \biggr]\mathbf{g}_F \\
&+(\xi-\frac{1}{6}) \lim_{z' \to z} Im \; \int \frac{d\omega dk_{\perp}}{(2\pi)^{3}} \biggr[
\frac{3}{2}k_\perp^2+\frac{3}{2}\frac{g_{11}}{g_{00}}\omega^2-\frac{3}{4}\Bigr(\partial_z^2 +\partial_{z'}^2\Bigr)+{\lambda_0+5\lambda_1\over 4}(\partial_z+\partial_{z'})
 \biggr]\mathbf{g}_F.
\end{split}
\end{eqnarray}
For the first line we find
\begin{eqnarray}\label{eq59}
\begin{split}
&\frac{1}{6} \biggr[
-\frac{3}{2}k_\perp^2-\frac{3}{2}\frac{g_{11}}{g_{00}}\omega^2 -\frac{3}{4}\Bigr(\partial_z^2 +\partial_{z'}^2-4\partial_z \partial_{z'}\Bigr)
+{\lambda_0+5\lambda_1\over 4}(\partial_z+\partial_{z'})
 \biggr]\mathbf{g}_F \\
&=\frac{1}{6}\biggr[
\frac{3}{2}Bz \frac{\omega^2}{\sqrt{b}} \frac{\cos(\sqrt{b}\beta)}{sin^2(\sqrt{b}l)}
-3(1-\gamma_0-\gamma_1)\frac{\sqrt{b}\cos(\sqrt{b}\beta)}{sin(\sqrt{b}l)}
+\frac{3l^2}{4}a\frac{\cos(\sqrt{b}\beta)\cos(\sqrt{b}l)}{sin^2(\sqrt{b}l)} \\
&\;\;\;\;\;\;\;\;\;\;\;\;\;\;\;\;\;\;\;\;\;\;\;\;\;\;\;\;\;\;\;\;\;\;+\frac{3l^2}{4}a\frac{\sin(\sqrt{b}\beta)}{sin(\sqrt{b}l)}
+6z\frac{\epsilon \sqrt{b} \cos(\sqrt{b}\beta)}{\sin(\sqrt{b}l)}
-\frac{9z}{4}\frac{a}{\sqrt{b}} \frac{\cos(\sqrt{b}l)}{\sin(\sqrt{b}l)}
\biggr] \\
&+\frac{1}{6}\biggr[
-\frac{3}{2}Bz \frac{\omega^2}{\sqrt{b}} \frac{\cos(\sqrt{b}\beta)}{sin^2(\sqrt{b}l)}
-\frac{3z}{4}\frac{a}{\sqrt{b}}\frac{\cos(\sqrt{b}\alpha)}{\sin(\sqrt{b}l)}
+3\epsilon \frac{ \sin(\sqrt{b}\alpha)}{\sin(\sqrt{b}l)}
-(\lambda_0+2\lambda_1)\frac{\cos(\sqrt{b}\alpha)}{\sin(\sqrt{b}l)}
\biggr],
\end{split}
\end{eqnarray}
and the second line of (\ref{eq58}) is found to be
\begin{eqnarray}\label{eq60}
\begin{split}
&\biggr[\frac{3}{2}k_\perp^2+\frac{3}{2}\frac{g_{11}}{g_{00}}\omega^2-\frac{3}{4}\Bigr(\partial_z^2 +\partial_{z'}^2\Bigr)+{\lambda_0+5\lambda_1\over 4}(\partial_z+\partial_{z'})
 \biggr]\mathbf{g}_F \\
&=\biggr[
-\frac{3}{2}Bz \frac{\omega^2}{\sqrt{b}} \frac{\cos(\sqrt{b}\beta)}{sin^2(\sqrt{b}l)}
-\frac{3z}{4}\frac{a}{\sqrt{b}} \frac{\cos(\sqrt{b}\beta)}{\sin(\sqrt{b}l)}
\biggr] \\
&+\biggr[
\frac{3}{2}Bz \frac{\omega^2}{\sqrt{b}} \frac{\cos(\sqrt{b}\alpha)}{sin^2(\sqrt{b}l)}
-\frac{3\lambda}{4}\frac{\sin(\sqrt{b}\alpha)}{\sin(\sqrt{b}l)}
+\frac{3z}{4}\frac{a}{\sqrt{b}}\frac{\cos(\sqrt{b}\alpha)}{\sin(\sqrt{b}l)}
-\frac{1}{4}(\lambda_0+5\lambda_1)\frac{\sin(\sqrt{b}\alpha)}{\sin(\sqrt{b}l)}
\biggr].
\end{split}
\end{eqnarray}

After doing the integrations we arrive at
\begin{eqnarray}\label{eq61}
\begin{split}
\langle T_{33} \rangle &
=\frac{3E_0}{l} \left( 1-2\gamma_1-\frac{2}{3}(2\lambda_0+\lambda_1)z+\frac{2}{3}(\lambda_0-\lambda_1)l\right)
+\frac{c_3}{(\Delta z)^4}     \\
&+(\xi-\frac{1}{6})\biggr[-\frac{1}{4\pi^2}( \lambda_0+2\lambda_1) A_4(\alpha)\biggr],
\end{split}
\end{eqnarray}
where
\begin{eqnarray}\label{eq62}
c_3=-\frac{3}{2\pi^2}\left(1-2\gamma_1-\frac{2}{3}(2\lambda_0+\lambda_1)z\right).
\end{eqnarray}
\subsection{consistency check}
As said before, the energy-momentum tensor has been found for a Casimir apparatus hovering in a weak static gravitational field described by Fermi coordinates \cite{Esposito,Napolitano,Bimonte}
\begin{eqnarray}\label{eq63}
ds^2=(1+2gz)dt^2-dx^2-dy^2-dz^2.
\end{eqnarray}
This spacetime is equivalent to the spacetime of the Rindler accelerated observer \cite{Bimonte1}. Therefore, it corresponds to the case $\gamma_0=\gamma_1=\lambda_1=0,\; \lambda_0=g$ in our calculations. They have found the following energy momentum tensor (see (4.5)-(4.7) in \cite{Esposito})
\begin{eqnarray}\label{eq64}
\begin{split}
&\langle T_{00} \rangle=\langle T^{(0)}_{00} \rangle+2gl \langle T^{(1)}_{00} \rangle+... \\
&\;\;\;\;\;\;\;\;\;\;\;\;\langle T^{(0)}_{00} \rangle=-\frac{\pi^2}{1440 l^4}=E_0, \\
&\;\;\;\;\;\;\;\;\;\;\;\;\langle T^{(1)}_{00} \rangle=...+E_0\frac{s-3\pi}{40}\frac{\cos(4s)}{\sin^4s}-... , \; s=\frac{\pi z}{l}\\
&\;\;\;\;\;\;\;\;\;\; \rightarrow \langle T_{00} \rangle=E_0\left(1+\frac{6}{5}ga-\frac{2}{5}gz) \right) , \\
&\langle T_{11} \rangle=\langle T_{22} \rangle=\langle T^{(0)}_{11} \rangle+2gl \langle T^{(1)}_{11} \rangle+... =-E_0+2ga\left(\frac{-E_0(\pi-2s)}{5\pi} \right) , \; s=\frac{\pi z}{l}\\
&\;\;\;\;\;\;\;=-E_0(1+\frac{2}{5}\lambda_0a-\frac{4}{5}\lambda_0z), \\
&\langle T_{33} \rangle=\langle T^{(0)}_{33} \rangle+2gl \langle T^{(1)}_{33} \rangle+... \\
&\;\;\;\;\;\;\;\;\;\;\;\;\langle T^{(0)}_{33} \rangle=3E_0, \\
&\;\;\;\;\;\;\;\;\;\;\;\;\langle T^{(1)}_{33} \rangle=E_0-\frac{2s}{\pi}E_0 \\
&\;\;\;\;\;\;\;\;\;\; \rightarrow \langle T_{33} \rangle=3E_0\left(1+\frac{2}{3}gl-\frac{4}{3}gz) \right). \\
\end{split}
\end{eqnarray}

Note that we have selected only the part of their result which is finite at $z=0$ and $z=l$. Thus, for instance, the only potentially finite term in $\langle T_{00} \rangle$ in that paper was the one containing the term $\cos(4s)/\sin^4s$ which has been demonstrated in $\langle T^{(1)}_{00} \rangle$ above. Also we have used $l$ instead of $a$ to show the separation between the plates. The $E_0$ is also the traditional Casimir energy of the flat spacetime. The result in equation (\ref{eq64}) equals our results for $\gamma_0=\gamma_1=\lambda_1=0,\; \lambda_0=g$.

Another consistency check concerns the covariant conservation of the energy momentum tensor.
Since the energy momentum tensor is diagonal and only dependent to $z$,  $\langle T^{\mu 0} \rangle_{;\mu}=\langle T^{\mu 1} \rangle_{;\mu}=\langle T^{\mu 2} \rangle_{;\mu}=0$. After a calculation for $\nu=3$, we find
\begin{eqnarray}\label{eq65}
\begin{split}
\langle T^{\mu 3} \rangle_{;\mu}=
&-\frac{1}{\pi^2}(2\lambda_0+\lambda_1)\frac{1}{\Delta z^4}
+\frac{c_4}{\Delta z^5},
\end{split}
\end{eqnarray}
in which
\begin{eqnarray}\label{eq66}
c_4=\frac{6}{\pi^2}( 1-6\gamma_1-\frac{4}{3}(\lambda_0+4\lambda_1)z ).
\end{eqnarray}

To obtain (\ref{eq65}) we have used the relation  $\partial_z A_4(\alpha)=2A_1(\alpha) $. As is well-known, the $\Delta z^{-n}$ terms in (\ref{eq65}) are the common effects of the point separation method which can be dropped away. Thus, the covariant conservation of the energy momentum tensor is guaranteed. Note also that the flat space limit can be easily checked out in view of (\ref{eq46}).
\subsection{The energy and the force}
The energy in a static spacetime is given by
\begin{eqnarray}\label{eq67}
E&=\int_{0}^{l} \sqrt{-g} \langle T^0_0 \rangle d^3x=S \int_{0}^{l} \left(1-\gamma_0+3\gamma_1+(3\lambda_1-\lambda_0)z\right) \langle T_{00} \rangle \nonumber \\
&=S(1+A+B\frac{l}{2})E_0+(\alpha-part)+(\xi-\frac{1}{6}) \left( \alpha-part \right),
\end{eqnarray}
where we have ignored the $\alpha$-part as it diverges at $z=0,l$. S is the area of plates.

Apparently, the first order correction $A=\gamma_0-\gamma_1$ has been appeared in the energy. This confirms the result recently found by author  \cite{borzooMPLA}. Although, in that work (see section 4 in \cite{borzooMPLA}), sufficient arguments were introduced for the appearance of first order corrections, the current direct calculation shows that undoubtedly the gravitational corrections for the Casimir energy and force of parallel plate geometry is many orders of magnitudes greater than what previously found in the literature and thus can be measured employing current precision of experiments.

The force by which the plates attract/repel each other is
\begin{eqnarray}\label{eq68}
\begin{split}
F&=-\frac{\partial E}{\partial l} =-S(1+A+\frac{2}{3}B\;l)\frac{\pi^2}{480 l^4}  \\
&=-S\left(1+\gamma_0+3\gamma_1+\frac{2}{3}(\lambda_0+2\lambda_1)\;l_p\right)\frac{\pi^2}{480 l_p^4},
\end{split}
\end{eqnarray}
where $l_p=\int_0^l \sqrt{-g_{33}}dz=l(1+\gamma_1+\frac{1}{2}\lambda_1l)$ is the proper distance between the plates. As a result, the change in the force by which the plates attract/repel each other depends on the sign of the first order correction $\gamma_0+3\gamma_1$. We give examples indicating this point later.

\section{conformal invariance and trace anomaly}
It can be shown that the trace of the stress-tensor vanishes for $\xi=1/6$, i.e. for the conformal coupling of the field. After some calculations and using equations (\ref{eq51}),(\ref{eq56}) and (\ref{eq61}) we find

\begin{eqnarray}\label{eq69}
\begin{split}
\langle T^\mu_\mu \rangle&=
g^{00}\langle T_{00}\rangle+g^{11}[2\langle T_{11}\rangle+\langle T_{33}\rangle]\\
&=0+\frac{1}{24\pi^2}(\xi-\frac{1}{6})\biggr[3(1-4\gamma_1-2\lambda z)A_1(\alpha)+6Bl^2A_2(\alpha)-6(2z^2-l^2)BA_3(\alpha) \\
&\;\;\;\;\;\;\;\;\;\;\;\;\;\;\;\;\;\;\;\;\;\;\;\;\;\;\;\;\;\;\;\;\;-(13\lambda_0+15\lambda_1)A_4(\alpha).
\biggr]
\end{split}
\end{eqnarray}
Therefore, the trace has an anomalous divergent part unless the field be conformally coupled.

Another feature of the obtained energy momentum tensor is related to the case the metric is conformal flat, i.e.
 \begin{eqnarray}\label{eq70}
\begin{split}
ds^2=(1+2\gamma_0+2\lambda_0 z)(dt^2-dx^2-dy^2-dz^2),  \;\;\;\; g^{00}=1-2\gamma_0-2\lambda_0z
\biggr].
\end{split}
\end{eqnarray}
In this case we have $\gamma_0=\gamma_1, \; \lambda_0=\lambda_1, \; A=0, \; B=0$ and the energy momentum tensor takes the form
\begin{eqnarray}\label{eq71}
\begin{split}
\langle T_{00} \rangle&=g^{00}\biggr[ \frac{E_0}{l}-\frac{1}{2\pi^2}\frac{1}{(\Delta z)^4}+\frac{1}{2\pi^2}(\xi-\frac{1}{6})A_1(2z-l) \biggr]+\frac{1}{2\pi^2}(\xi-\frac{1}{6})[-3\lambda_0A_4(2z-l)]  \\
&=g^{00}\langle T_{00} \rangle^{flat}-(\xi-\frac{1}{6})\frac{3\lambda_0}{2\pi^2}A_4(2z-l), \\
\langle T_{11} \rangle&=\langle T_{22} \rangle    \\
&=g^{00}\biggr[ -\frac{E_0}{l}+\frac{1}{2\pi^2}\frac{1}{(\Delta z)^4}-\frac{1}{2\pi^2}(\xi-\frac{1}{6})A_1(2z-l) \biggr]+\frac{1}{2\pi^2}(\xi-\frac{1}{6})[3\lambda_0A_4(2z-l)],  \\
&=g^{00}\langle T_{11} \rangle^{flat}+(\xi-\frac{1}{6})\frac{3\lambda_0}{2\pi^2}A_4(2z-l), \\
\langle T_{33} \rangle&=g^{00}\biggr[ 3\frac{E_0}{l}-\frac{3}{2\pi^2}\frac{1}{(\Delta z)^4} \biggr]+\frac{1}{2\pi^2}(\xi-\frac{1}{6})[-3\lambda_0A_4(2z-l)],  \\
&=g^{00}\langle T_{33} \rangle^{flat}-(\xi-\frac{1}{6})\frac{3\lambda_0}{2\pi^2}A_4(2z-l).
\end{split}
\end{eqnarray}
This is the reminiscence of the already known relation in the literature. If a metric undergoes a conformal transformation $\bar{g}_{\mu\nu}=\Omega^2(x)g_{\mu\nu}$, the new (renormalized) energy momentum tensor is given by (see (6.134) in \cite{Birrell})
\begin{eqnarray}\label{eq72}
\begin{split}
\langle T_\mu^\nu[\bar{g}_{\mu\nu}] \rangle_{ren.}=&(\frac{g}{\bar{g}})^{\frac{1}{2}}\langle T_\mu^\nu[g_{\mu\nu}] \rangle_{ren.}  \\
&+\frac{1}{12}\biggr[ (\Omega^{-3}\Omega_{;\theta\mu}-2\Omega^{-4}\Omega_{;\theta}\Omega_{;\mu})g^{\theta\nu} +\delta_\mu^\nu g^{\rho\sigma}(\frac{3}{2}\Omega^{-4}\Omega_{;\rho}\Omega_{;\sigma}-\Omega^{-3}\Omega_{;\rho\sigma})g^{\theta\nu}\biggr].
\end{split}
\end{eqnarray}
Putting $\Omega^2=g_{00}=1+2\gamma_0+2\lambda_0z$ we see that
\begin{eqnarray}\label{eq73}
\langle T_{\mu\nu} \rangle_{ren.}&=g^{00}\langle T_{\mu\nu} \rangle^{flat}_{ren.}+O(\lambda^2),
\end{eqnarray}
which differs from what we found in (\ref{eq71}) by a factor of $(\xi-\frac{1}{6})\frac{3\lambda_0}{2\pi^2}A_4(2z-l)$. This difference is related to the fact that equation (\ref{eq72}) has been derived for quantum field theory in curved spacetime without boundaries while our result is obtained in the presence of boundary. An improved form of (\ref{eq72}) and some other calculations related to quantum field theory in curved spacetime under the influence of boundaries will be published elsewhere.

Again, equation (\ref{eq73}) inspects equation (\ref{eq71}) in the case of conformal coupling, i.e. $\xi=\frac{1}{6}$. In other words, for the case of conformal triviality, the divergent part $A_4(2z-l)$ disappears. 

\section{examples}
\subsection{The Kerr spacetime}
The Casimir effect in Kerr spacetime has been studied previously \cite{SorgeNew,Bezerra,Toshmatov}. The metric of a slowly rotating source adapted to the Casimir plates, measured by a zero angular momentum observer (ZAMO), is given by \cite{SorgeNew}
\begin{eqnarray}\label{eq74}
\begin{split}
ds^2=(1+2b\Phi_0+2b\eta z)dt^2-(1-2\Phi_0-2\eta z)(dx^2+dy^2+dz^2),
\end{split}
\end{eqnarray}
in which
\begin{eqnarray}\label{eq75}
\begin{split}
&b=1-2a\Omega,\;\;\; \Omega=\frac{d\phi}{dt}, \;\;\; a=J/M\\
&\Phi_0=-\frac{Gm}{c^2R}, \; \eta=\frac{Gm}{c^2R^2}.
\end{split}
\end{eqnarray}
For more general observers see \cite{SorgeNew}. Note that $m$ is the mass of the source and the apparatus is located at distance $R$ from the center of the source. $a$ is the angular momentum per mass.

By comparing (\ref{eq74}) with (\ref{eq1}) we see that $\gamma_0=b\Phi_0, \; \gamma_1=-\Phi_0$. Thus, in view of equation (\ref{eq68}), we have $\gamma_0+3\gamma_1=-2\Phi_0(1+a\Omega)>0$ because for zero angular momentum observers $\Omega\approx\frac{2Ma}{R^3}$ in the far field limit. As a result, the magnitude of the force between the plates increases. This is also the case for the Schwarzschild metric as it corresponds to $\Omega=0$ which does not alter the sign of $\gamma_0+3\gamma_1$.

\subsection{Extended theories of gravity(ETG)}
The metric (\ref{eq1}) is applicable also for the case of the extended theories of gravity. The possible impact of such theories on the Casimir energy has been studied in \cite{Lambiase} where they have found the related metric to be of the form
\begin{eqnarray}\label{eq76}
\begin{array}{ll}
g_{00}(\mathbf{x})\simeq  1+2\,\Phi_{0}(R) +2\,\Lambda(R)\,z
\\
g_{ij}(\mathbf{x})\simeq -1+2\,\Psi_{0}(R) +2\,\Sigma(R)\,z \;,
\end{array}
\end{eqnarray}
where
\begin{eqnarray}\label{eq77}
\begin{array}{ll}
\Phi_{0}(R)\,=\,-\frac{GM}{R}\biggl[1+g(\xi,\eta)\,e^{-m_+ R}+\biggl(\frac{1}{3}-g(\xi,\eta)\biggr)\,e^{-m_- R}-\frac{4}{3}\,e^{-m_Y R}\biggr]
\\
\Lambda(R)\,=\,\frac{GM}{R^2}\biggl[1+g(\xi,\eta)\ \bigl(1+m_+ R\bigl)\,e^{-m_+ R}+\biggl(\frac{1}{3}-g(\xi,\eta)\biggr)\bigl(1+m_- R\bigl) \,e^{-m_- R}+\\
\qquad\qquad\qquad\qquad\qquad\qquad\qquad\qquad\qquad\qquad\qquad\qquad\qquad-\frac{4}{3}\bigl(1+m_Y R\bigl) \,e^{-m_Y R}\biggr]
\\
\Psi_{0}(R)\,=\,-\frac{GM}{R}\biggl[1-g(\xi,\eta)\,e^{-m_+ R}-\biggl(\frac{1}{3}-g(\xi,\eta)\biggr)\,e^{-m_- R}-\frac{2}{3}\,e^{-m_Y R}\biggr]
\\
\Sigma(R)\,=\,\frac{GM}{R^2}\biggl[1-g(\xi,\eta)\bigl(1+m_+ R\bigl)\,e^{-m_+ R}-\biggl(\frac{1}{3}-g(\xi,\eta)\biggr)\bigl(1+m_- R\bigl)\,e^{-m_- R}+\\
\qquad\qquad\qquad\qquad\qquad\qquad\qquad\qquad\qquad\qquad\qquad\qquad\qquad-\frac{2}{3}\bigl(1+m_Y R\bigl)\,e^{-m_Y R}\biggr].
\end{array}
\end{eqnarray}
The parameters $m_+,m_-,m_Y$ and $g(\xi,\eta)$ have been defined in \cite{Lambiase}. By now, it is sufficient to know that the main term in the above equations is the Newtonian potential $GM/R$ and extra terms in the brackets are corrections due to ETGs. Therefore, $\gamma_0=\Phi_{0}, \; \lambda_0=\Lambda,\; \gamma_1=-\Psi_0,\; \lambda_1=-\Sigma$ and the corresponding Casimir force is given by
\begin{eqnarray}\label{eq78}
F=-S\biggl[1+\Phi_{0}-3\Psi_0 +\frac{2}{3}(\Lambda-2\Sigma) l_P\biggr]\frac{\pi^2\,}{480\,l^4_p}\,,
\end{eqnarray}
which shows an increase in the magnitude of the force.
\subsection{Horava-Lifshitz gravity}
Modifications of the Casimir energy by the Horava-Lifshitz theory of gravity has been studied in \cite{Muniz1}. They have found a static black hole solution as follows (see eq.(8) in \cite{Muniz1})
\begin{eqnarray}\label{eq79}
ds^2=\left(1-\frac{2M}{R}+\frac{2M^2}{\omega R^4}\right)dt^2-\left(1+\frac{2M}{R}-\frac{M^2}{2\omega R^4}\right)(dr^2+r^2d\Omega^2),
\end{eqnarray}

which in turn gives $\gamma_0=-\frac{M}{R}=-\gamma_1,\; \lambda_0=\frac{M^2}{\omega R^4}=-\frac{1}{4}\lambda_1$. Thus, the magnitude of the Casimir force increases as $\gamma_0+3\gamma_1=2\frac{M}{R}>0$.
\section{concluding remarks}
In this paper, we found the curved spacetime analogue of the energy-momentum tensor for Casimir effect of parallel plates first found by Brown and Maclay \cite{Brown} using the point splitting method. We extended in detail the calculations of the energy momentum tensor of the arbitrarily coupled scalar field confined in between the Casimir plates to the metric given by equation (\ref{eq1}). As we shown, the metric (\ref{eq1}) covers all previously considered static weak gravitational fields for which the Casimir energy and force has been calculated in the literature. The explicit structure of divergencies was determined and the regularized stress tensor was obtained in equations (\ref{eq51}),(\ref{eq56}) and (\ref{eq61}). Consistency with the previous results in the literature were done in subsection G and section IV.

We found sufficient conditions according to which the force and energy decreases/increases. Also we proved directly (eqs.(67-68), (75) and (78-79) ) that the leading order corrections to both the Casimir energy and force in flat spacetime is $\frac{Gm}{c^2R}$ rather than $\frac{Gm}{c^2R^2}$ where previously found in the literature. Note that $\frac{Gm}{c^2R}$ is dimensionless and some order of magnitudes greater than $\frac{Gm}{c^2R^2}$ as $R$ can be thought as radius of the source of gravity, say $R=6.38 \times 10^6$ meters. We found the energy momentum tensor in the case of conformal coupling of the field and shown the consistency of the results. Some previous studies in the literature were analysed and corrected through examples.

\pagebreak
\appendix
\section{Some important computations}
To be concise, we use equation (\ref{eq17a}) in the form
\begin{eqnarray}\label{A1}
\mathbf{g}_F= \frac{1-\gamma_0-\gamma_1-\epsilon(z+z')}{2\sqrt{b} \sin \sqrt{b}\left(l+\frac{a}{4b}l^2\right)}\biggr[\cos (S_1+S_2)-\cos(S_1-S_2)\biggr],
\end{eqnarray}

in which
\begin{eqnarray}\label{A2}
\begin{split}
&S_1=\sqrt{b}z+\frac{a}{4\sqrt{b}}z^2, \\
&S_2=\sqrt{b}(\acute{z}-l)+\frac{a}{4\sqrt{b}}(\acute{z}^2-l^2).
\end{split}
\end{eqnarray}
Using the above equations we find up to second order perturbations
\begin{eqnarray}\label{A3}
\partial_z \mathbf{g}_F&=&\frac{-\epsilon}{2\sqrt{b}\sin(\sqrt{b}l)}\left(\cos(\sqrt{b}\alpha)-\cos(\sqrt{b}\beta) \right) -\frac{az}{4b\sin(\sqrt{b}l)}\left(\sin(\sqrt{b}\alpha)-\sin(\sqrt{b}\beta) \right) \nonumber \\
&+&\frac{\sqrt{b}}{2R}(1-\epsilon (z+z')) \left(\sin(S_1+S_2)-\sin(S_1-S_2) \right),
\end{eqnarray}
in which $R=-(1+\gamma_0+\gamma_1)\sqrt{b} \sin(\sqrt{b}l+\frac{a}{4\sqrt{b}}l^2)$ and equation (\ref{eq39}) has been used. In a same way we find
\begin{eqnarray}\label{A4}
 \partial_{z'} \mathbf{g}_F&=&\frac{-\epsilon}{2\sqrt{b}\sin(\sqrt{b}l)}\left(\cos(\sqrt{b}\alpha)-\cos(\sqrt{b}\beta) \right)
-\frac{az'}{4b\sin(\sqrt{b}l)}\left(\sin(\sqrt{b}\alpha)+\sin(\sqrt{b}\beta) \right) \nonumber \\
&+&\frac{\sqrt{b}}{2R}(1-\epsilon (z+z'))
\left( \sin(S_1+S_2)+\sin(S_1-S_2) \right),
\end{eqnarray}
\begin{eqnarray}\label{A5}
\begin{split}
\partial_{z'} \partial_{z} \mathbf{g}_F&=\epsilon \frac{\sin(\sqrt{b}\alpha)}{\sin(\sqrt{b}l)}
-\frac{a(z+z')}{4\sqrt{b}\sin(\sqrt{b}l)}\left(\cos(\sqrt{b}\alpha)+\cos(\sqrt{b}\beta) \right) \\
&\;\;\;\;\;\;\;\;\;\;\;\;\;\;\;\;\;\;\;\;\;\;\;\;\;\;\;\;\;\;\;\;\;\;\;\;\;\;+\frac{b}{2R}(1-\epsilon (z+z'))
\left( \cos(S_1+S_2)+\cos(S_1-S_2) \right),
\end{split}
\end{eqnarray}
\begin{eqnarray}\label{A6}
\partial^2_z \mathbf{g}_F&=&\frac{\lambda}{2\sin(\sqrt{b}l)}\left(\sin(\sqrt{b}\alpha)-\sin(\sqrt{b}\beta) \right) -\frac{az}{2\sqrt{b}\sin(\sqrt{b}l)}\left(\cos(\sqrt{b}\alpha)-\cos(\sqrt{b}\beta) \right) \nonumber \\
&+&\frac{b}{2R}(1-\epsilon (z+z')) \left(\cos(S_1+S_2)-\cos(S_1-S_2) \right),
\end{eqnarray}
\begin{eqnarray}\label{A7}
\partial^2_z \mathbf{g}_F&=&\frac{\lambda}{2\sin(\sqrt{b}l)}\left(\sin(\sqrt{b}\alpha)+\sin(\sqrt{b}\beta) \right) -\frac{az'}{2\sqrt{b}\sin(\sqrt{b}l)}\left(\cos(\sqrt{b}\alpha)-\cos(\sqrt{b}\beta) \right) \nonumber \\
&+&\frac{b}{2R}(1-\epsilon (z+z')) \left(\cos(S_1+S_2)-\cos(S_1-S_2) \right),
\end{eqnarray}
The last term in equations (\ref{A5})-(\ref{A7}) should be further simplified to be applicable practically. After some calculations we find that
\begin{eqnarray}\label{A8}
\begin{split}
&\frac{1}{R}(1-\epsilon (z+z')) \cos(S_1+S_2)=M, \\
&\frac{1}{R}(1-\epsilon (z+z')) \cos(S_1-S_2)=N,
\end{split}
\end{eqnarray}
where $M$ and $N$ were defined in equation (\ref{eq35a}) and (\ref{eq35b}). It is stressed again that all the above relations have been approximated up to second order perturbations using equation (\ref{eq39}).
Putting equations (\ref{A8}) back into (\ref{A5})-(\ref{A7}) will produce (\ref{eq35a}),(\ref{eq36}) and (\ref{eq37}).

\section{Wick rotation for intergrations}
Suppose we tend to compute, for instance, the integral
\begin{eqnarray}\label{B1}
Y=\int \frac{d\omega d^2k_\perp}{(2\pi)^3} \frac{a\cos(\sqrt{b}l)}{\sqrt{b}\sin(\sqrt{b}l)}.
\end{eqnarray}
In this appendix $b=\omega^2-k_\perp^2$. If $b=(1-2A)\omega^2-k_\perp^2$ which is the case for our problem in this paper, in any case, the variable change $\omega'=(1-A)\omega$ will recast the integration to $b=\omega^2-k_\perp^2$. The Wick rotation is achieved by sending
\begin{eqnarray}\label{B2}
\omega\rightarrow i\kappa cos\theta, \; k_\perp \rightarrow \kappa \sin\theta \; \Rightarrow \sqrt{b} \rightarrow i\kappa,\; \Rightarrow \frac{d\omega d^2k_\perp}{(2\pi)^3}=\frac{2i}{(2\pi)^2}\kappa^2 d\kappa \sin\theta d\theta.
\end{eqnarray}
Since $a=-2B\omega^2$ we find
\begin{eqnarray}\label{B4}
Y=\frac{-Bi}{3\pi^2}\int \frac{\kappa^3\cosh(\kappa l)}{\sinh(\kappa l)}d\kappa=\frac{-Bi}{3\pi^2} \lim_{z \to z'}A_1(\beta),
\end{eqnarray}
in which we have defined
\begin{subequations}
\begin{align}
&A_1(u)=\int_{0}^{\infty} \frac{\kappa^3 \cosh(\kappa u)}{\sinh\kappa l} d\kappa, \label{B6a} \\
&A_2(u)=\int_{0}^{\infty} \frac{\kappa^4 \cosh^2(\kappa u)}{\sinh^2\kappa l} d\kappa, \label{B6b} \\
&A_3(u)=\int_{0}^{\infty} \frac{\kappa^4 \sinh(\kappa u)}{\sinh\kappa l} d\kappa, \label{B6c}\\
&A_4(u)=\int_{0}^{\infty} \frac{\kappa^2 \sinh(\kappa u)}{\sinh\kappa l} d\kappa. \label{B6d}
\end{align}
\end{subequations}
Other integrations which are needed in the paper can be find as follows:

\begin{subequations}
\begin{align}
&\int \frac{d\omega dk_{\perp}}{(2\pi)^{3}} \frac{\omega^2}{\sqrt{b}}\frac{\cos(\sqrt{b}u)}{\sin(\sqrt{b}l)}=\frac{1}{3}\frac{2i}{(2\pi)^2} A_1(u), \\
&\int \frac{d\omega dk_{\perp}}{(2\pi)^{3}} \frac{a\;\omega^2}{b}\frac{\cos^2(\sqrt{b}u)}{\sin^2(\sqrt{b}l)}=\frac{-2B}{5}\frac{2i}{(2\pi)^2} A_2(u), \\
&\int \frac{d\omega dk_{\perp}}{(2\pi)^{3}} \frac{a\;\omega^2}{b}\frac{\sin(\sqrt{b}u)}{\sin(\sqrt{b}l)}=\frac{2B}{5}\frac{2i}{(2\pi)^2} A_3(u), \\
&\int \frac{d\omega dk_{\perp}}{(2\pi)^{3}}
\frac{\epsilon \; \omega^2}{\sqrt{b}}\frac{\cos(\sqrt{b}u)}{\sin(\sqrt{b}l)}
=\frac{1}{2}(\frac{\lambda}{3}-\frac{B}{5})\frac{2i}{(2\pi)^2} A_1(u), \\
&\int \frac{d\omega dk_{\perp}}{(2\pi)^{3}} \frac{k_\perp^2}{\sqrt{b}}\frac{\cos(\sqrt{b}u)}{\sin(\sqrt{b}l)}=-\frac{2}{3}\frac{2i}{(2\pi)^2} A_1(u), \\
&\int \frac{d\omega dk_{\perp}}{(2\pi)^{3}}
\frac{k_\perp^2 \; a}{b}\frac{\cos^2(\sqrt{b}u)}{\sin^2(\sqrt{b}l)}=\frac{4B}{15}\frac{2i}{(2\pi)^2} A_2(u), \\
&\int \frac{d\omega dk_{\perp}}{(2\pi)^{3}}
\frac{k_\perp^2 \; a}{b}\frac{\sin(\sqrt{b}u)}{\sin(\sqrt{b}l)}=-\frac{4B}{15}\frac{2i}{(2\pi)^2} A_3(u), \\
&\int \frac{d\omega dk_{\perp}}{(2\pi)^{3}}
\frac{\epsilon \; k_\perp^2 }{\sqrt{b}}\frac{\cos(\sqrt{b}u)}{\sin(\sqrt{b}l)}=-(\frac{\lambda}{3}-\frac{B}{15})\frac{2i}{(2\pi)^2} A_1(u),  \\
&\int \frac{d\omega dk_{\perp}}{(2\pi)^{3}}
\epsilon\frac{\sin(\sqrt{b}u)}{\sin(\sqrt{b}l)}=(\frac{\lambda}{2}-\frac{B}{6})\frac{2i}{(2\pi)^2} A_4(u), \\
&\int \frac{d\omega dk_{\perp}}{(2\pi)^{3}} \frac{a}{\sqrt{b}}\frac{\cos(\sqrt{b}u)}{\sin(\sqrt{b}l)}=-\frac{2B}{3}\frac{2i}{(2\pi)^2} A_1(u) , \\
&\int \frac{d\omega dk_{\perp}}{(2\pi)^{3}} \sqrt{b}\frac{\cos(\sqrt{b}u)}{\sin(\sqrt{b}l)}=\frac{2i}{(2\pi)^2} A_1(u), \\
&\int \frac{d\omega dk_{\perp}}{(2\pi)^{3}} a\frac{\cos^2(\sqrt{b}u)}{\sin^2(\sqrt{b}l)}=\frac{-2B}{3}\frac{2i}{(2\pi)^2} A_2(u), \\
&\int \frac{d\omega dk_{\perp}}{(2\pi)^{3}} a\frac{\sin(\sqrt{b}u)}{\sin(\sqrt{b}l)}=\frac{2B}{3}\frac{2i}{(2\pi)^2} A_3(u), \\
&\int \frac{d\omega dk_{\perp}}{(2\pi)^{3}}
\epsilon \; \sqrt{b} \frac{\cos(\sqrt{b}u)}{\sin(\sqrt{b}l)}=(\frac{\lambda}{2}-\frac{B}{6})\frac{2i}{(2\pi)^2} A_1(u), \\
&\int \frac{d\omega dk_{\perp}}{(2\pi)^{3}} \frac{\sin(\sqrt{b}u)}{\sin(\sqrt{b}l)}=\frac{2i}{(2\pi)^2} A_4(u), \\
&\int \frac{d\omega dk_{\perp}}{(2\pi)^{3}} \sqrt{b}\frac{\cos(\sqrt{b}u)}{\sin(\sqrt{b}l)}=\frac{2i}{(2\pi)^2} A_1(u).
\end{align}
\end{subequations}
In our calculations we need either $u=\alpha=z+z'-l$ or $u=\beta=z-z'+l=\Delta z+l$. Thus, we find the following results

\begin{subequations}
\begin{align}
\lim_{z \to z'}A_1(\beta)&=\lim_{z \to z'}\int_{0}^{\infty} \frac{\kappa^3 \cosh(\kappa (\Delta z+l))}{\sinh\kappa l} d\kappa \nonumber \\
&=\lim_{z \to z'}2\int_{0}^{\infty} \frac{\kappa^3 \cosh(\kappa \Delta z)}{e^{2\kappa l}-1}+\lim_{z \to z'}\int_{0}^{\infty} \kappa^3e^{k\Delta z}=
\frac{\pi^4}{120l^4}+\lim_{\Delta z \to 0}\frac{6}{(\Delta z)^4}, \label{B6a} \\
\lim_{z \to z'} A_2(\beta)&=\lim_{z \to z'}\int_{0}^{\infty} \frac{\kappa^4 \cosh(\kappa (\Delta z+l))\cosh(\kappa l)}{(e^{2\kappa l}-1)^2} d\kappa \nonumber \\
&= \lim_{z \to z'} \Biggr[
4\int_{0}^{\infty} \frac{\kappa^4 }{e^{2\kappa l}-1}
+4\int_{0}^{\infty} \frac{\kappa^4 }{(e^{2\kappa l}-1)^2}
+\int_{0}^{\infty} \kappa^4e^{2k\Delta z}
\Biggr] \nonumber \\
&=\frac{\pi^4}{30l^5}-\lim_{\Delta z \to 0}\frac{24}{(\Delta z)^5},      \label{B6b} \\
\lim_{z \to z'}A_3(\beta)&=\lim_{z \to z'} \int_{0}^{\infty} \frac{\kappa^4 \sinh(\kappa (\Delta z+l))}{\sinh\kappa l} d\kappa=-\lim_{\Delta z \to 0}\frac{4!}{(\Delta z)^5}, \label{B6c}\\
\lim_{z \to z'}A_4(\beta)&=\lim_{ z \to z'} \int_{0}^{\infty} \frac{\kappa^2 \sinh(\kappa (\Delta z+l))}{\sinh\kappa l} d\kappa=-\lim_{\Delta z \to 0}\frac{2}{(\Delta z)^3} \label{B6d},
\end{align}
\end{subequations}
in which we have used (\ref{eq45}) for obtaining $A_1(\beta),A_2(\beta),A_3(\beta),A_4(\beta)$.

To obtain similar results for $\alpha$-part, we use suitable decomposition of partial fractions and the relations 3.524(5) and 3.423(2) in \cite{gradshtyn}. Then it is found that
\begin{subequations}
\begin{align}
\lim_{z \to z'}A_1(\alpha)&=\int_{0}^{\infty} \frac{\kappa^3 \cosh(\kappa (2z-l))}{\sinh\kappa l} d\kappa
=\frac{3}{8l^4}\biggr[ \zeta(4,1-\frac{z}{l})+\zeta(4,\frac{z}{l})\biggr], \label{B8a} \\
\lim_{z \to z'} A_2(\alpha)&=\int_{0}^{\infty} \frac{\kappa^4 \cosh(\kappa (2z-l))\cosh(\kappa l)}{\sinh^2\kappa l} d\kappa \nonumber \\
&=2\int_{0}^{\infty} \frac{\kappa^4 [e^{2\kappa z }+e^{2\kappa (l-z) }]}{(e^{2\kappa l}-1)^2} d\kappa
+\int_{0}^{\infty} \frac{\kappa^4 [e^{2\kappa z }+e^{2\kappa (l-z) }]}{e^{2\kappa l}-1} d\kappa  \nonumber \\
&=\frac{3}{4l^5} \biggr[ 2\zeta(4,2-\frac{z}{l})-2(1-\frac{z}{l})\zeta(5,2-\frac{z}{l})  \nonumber \\
&+2\zeta(4,1+\frac{z}{l})-\frac{2z}{l}\zeta(5,1+\frac{z}{l})+\zeta(5,1-\frac{z}{l})+\zeta(5,\frac{z}{l})
\biggr],
\label{B8b} \\
\lim_{z \to z'}A_3(\alpha)&=\int_{0}^{\infty} \frac{\kappa^4 \sinh(\kappa (2z-l))}{\sinh\kappa l} d\kappa
=\frac{3}{4l^5}\biggr[ \zeta(5,1-\frac{z}{l})-\zeta(5,\frac{z}{l})\biggr], \label{B8c}\\
\lim_{z \to z'}A_4(\alpha)&=\int_{0}^{\infty} \frac{\kappa^2 \sinh(\kappa (2z-l))}{\sinh\kappa l} d\kappa
=\frac{1}{4l^3}\biggr[ \zeta(3,1-\frac{z}{l})-\zeta(3,\frac{z}{l})\biggr]  \label{B8d},
\end{align}
\end{subequations}
where the following relation has been used to obtain $A_2(\beta)$ (see 3.423(2) in \cite{gradshtyn,odintsov2})
\begin{eqnarray}\label{B5}
\begin{split}
&\int_{0}^{\infty}\frac{x^\nu e^{-\mu x}}{(e^x-1)^2}dx=\Gamma(\nu+1)\biggr[ \zeta(\nu,\mu+2)-(\mu+1)\zeta(\nu+1,\mu+2) \biggr], \\
&\;\;\;\;\;\;\;\;\;\;\;\;\;\;\;\;\;\;\;\;\;\;\;\;\;\;\;\;\;\;\;\;\;\;\;\;\;Re\;\mu>-2,\; Re\;\nu>2.
\end{split}
\end{eqnarray}

\section *{Acknowledgments}
 Borzoo Nazari would like to thank University of Tehran for supporting this research.


\begin{thebibliography}{30}
\bibitem{Brown}  L. S. Brown and G. J. Maclay, Phys. Rev. 184,  1272 (1969).
\bibitem{Hawking} S. W. Hawking, Nature 248, 30–31, (1974). https://doi.org/10.1038/248030a0
\bibitem{Parker} L. Parker, The Physical Review 5, 183 (1969);L. Parker, The Physical Review 2 ,3 (1972).
\bibitem{Birrell} N.D. Birrell, and P.C.W. Davies, Quantum field theory in curved spacetime, Cambridge universsity press, 1982.
\bibitem{DeWitt} B. S. DeWitt, Phys. Rep. 19, 295-357 (1975).
\bibitem{Milton} K. A. Milton , P. Parashar, K. V. Shajesh, and J. Wagner, J. Phys. A 40,10935-10943,(2007).
\bibitem{Fulling} S. A. Fulling,  K. A. Milton, P. Parashar, A. Romeo, K.V. Shajesh, J. Wagner  Phys. Rev. D 76, 025004 (2007).
\bibitem{Milton2} K. A. Milton, K. V. Shajesh, S. A. Fulling, Prachi Parashar, Phys. Rev. D 89, 064027 (2014).
\bibitem{Shajesh} K V Shajesh, Kimball A Milton, Prachi Parashar and Jeffrey A Wagner, J. Phys. A: Math. Theor. 41, 164058 (2008).
\bibitem{Calloni} E. Calloni, L.Di Fiore, G. Esposito, L. Milano and L. Rosa, Phys. Lett. A 297, 328-33 (2002).
\bibitem{NouriNazari} M. Nouri-zonoz, Borzoo Nazari, Phys. Rev. D 82, 044047 (2010) .

\bibitem{NouriNazari2} Borzoo Nazari, M. Nouri-zonoz , Phys. Rev. D 85, 044060 (2012).
\bibitem{Saharian} Aram A. Saharian, Phys. Rev. D 70, 064070 (2004).
\bibitem{Bezerra} V.B. Bezerra, H.F.Mota, C.R. Muniz,  Phys.Rev.D 89,044015 (2014).
\bibitem{SorgeNew} F. Sorge, Phys. Rev. D 90, 084050 (2014).
\bibitem{Muniz1} C.R. Muniz, V.B. Bezerra, M.S. Cunha, Phys. Rev. D 88, 104035(2013).
\bibitem{Muniz2} C.R. Muniz, V.B. Bezerra, M.S. Cunha, Annals of Physics 359, 55-63 (2015) .
\bibitem{Sorge} F. Sorge, Class. Quantum. Grav. 22, 5109-5119 (2005).
\bibitem{Sorge2019} F. Sorge, Class. Quantum Grav. 36, 235006 (2019).
\bibitem{Lima} A.P.C.M Lima, G. Alencar, C.R. Muniz and R.R. Landim, JCAP 07, 011 (2019).
\bibitem{Blasone} Massimo Blasone, Gaetano Lambiase, Luciano Petruzziello, Antonio Stabile, Eur. Phys. J. C  78, 976 (2018).
\bibitem{Buoninfante} Luca Buoninfante, Gaetano Lambiase, Luciano Petruzziello, Antonio Stabile, Eur. Phys. J. C 79, 41 (2019).
\bibitem{Lambiase} G. Lambiase, A. Stabile and An. Stabile, Phys. Rev.D 95, 084019 (2017).
\bibitem{BorzooEPJC} Borzoo Nazari, Eur. Phys. J. C 75, 501 (2015).
\bibitem{BorzooCQG}  Borzoo Nazari, Class. Quantum Grav. 37, 135014 (2020).
\bibitem{Teo2} L.P. Teo   JHEP 10, 019 (2010); P. Wongjun, Eur. Phys. J. C 75, 6 (2015); J. Lorenzen and D. Martelli  JHEP 07, 001 (2015); A. Edery and V. Marachevsky, JHEP 12, 035 (2008).
\bibitem{Saharian1} Aram A. Saharian, Phys. Rev. D 69, 085005 (2004).
\bibitem{Saharian2} A. A. Saharian, Int. J. Mod. Phys. A 26, 3833–3844 (2011).
\bibitem{odintsov1}  E. Elizalde, S. D. Odintsov and A. A. Saharian, Phys. Rev. D 83, 105023 (2011);
                     E. Elizalde, S. D. Odintsov and A. A. Saharian, Phys. Rev. D 79, 065023 (2009).
\bibitem{Bimonte} G. Bimonte, E. Calloni, G. Esposito, and L. Rosa, Phys. Rev. D 74, 085011 (2006) ; Erratum ibid. D 75, 049904 (2007); Erratum ibid. D 75, 089901 (2007); Erratum ibid. D 77, 109903 (2008).
\bibitem{Bimonte1} G. Bimonte, G. Esposito, and L. Rosa, Phys. Rev. D 78, 024010 (2008) .
\bibitem{Bimonte2} G. Bimonte, E. Calloni, G. Esposito and L. Rosa, Phys. Rev. D 76, 024010 (2007) .
\bibitem{Esposito} G. Esposito, G. M. Napolitano,  L Rosa, Phys.Rev.D 77, 105011 (2008).
\bibitem{Napolitano} G. M. Napolitano, G. Esposito, L Rosa, Phys.Rev.D 78, 107701 (2008).
\bibitem{Misner} C.W. Misner, K. S. Thorne, J. H. Wheeler,  Gravitation, W.H. Freeman and company, 1973.
\bibitem{Christensen} S. M. Christensent, Phys.Rev. D 14, 2490 (1976).
\bibitem{Wylie} C. Ray Wylie and L. C. Barrett, Advanced engineering mathematics, 6th Ed., McGraw-Hill Inc., 1995, page 181.
\bibitem{Hassani} S. Hassani, Mathematical Physics:A Modern Introduction to Its Foundations, Second Edition, Springer, 2013, see Equ.(20.23).
\bibitem{Toshmatov} B. Toshmatov , Z. Stuchlik,  B. Ahmedov, Mod. Phys. Lett. A 32, 1775001 (2017).
\bibitem{borzooMPLA} Borzoo Nazari, Quasi-local stress-tensor formalism and the Casimir effect, Mod. Phys. Lett. A 37 2250160 (2022).
    https://doi.org/10.1142/S0217732322501607

\bibitem{odintsov2} E. Elizalde, S. D. Odintsov, A. Romeo, A. A. Bytsenko and S. Zerbini, Zeta Function Regularization Techniques with Applications (World Scientific Publishing, 1994), doi:10.1142/2065.
\bibitem{gradshtyn} I.S. Gradshteyn, I.M. Ryzhik, Table of Integrals, Series and Products, 7th edition, Elsevier Inc., 2007.


\end{thebibliography}
\end{document}